\documentclass[preprint]{JHEP3} 

\usepackage{epsfig}
\usepackage{amsmath}
\usepackage{amssymb,amsfonts}
\usepackage{graphicx}
\usepackage{graphics}
\usepackage{amsthm}
\usepackage{graphicx}
\usepackage{multirow}

\newcommand{\cO}{{\cal O}}
\newcommand{\cA}{{\cal A}}
\newcommand{\cAb}{{\overline{\cal A}}}
\newcommand{\cF}{{\cal F}}
\newcommand{\cFb}{{\overline{\cal F}}}
\newcommand{\cD}{{\cal D}}
\newcommand{\cDb}{{\overline{\cal D}}}
\newcommand{\cQ}{{\cal Q}}
\newcommand{\cU}{{\cal U}}
\newcommand{\cN}{{\cal N}}
 
\newcommand{\Tr}{{\rm Tr\;}}
\newcommand{\etab}{{\overline{\eta}}}
\newcommand{\psib}{{\overline{\psi}}}
\newcommand{\phib}{{\overline{\phi}}}

\newcommand{\br}{{\boldsymbol r}}
\newcommand{\bR}{{\boldsymbol R}}
\newcommand{\hatbe}{\widehat{\boldsymbol e}}

\newcommand{\hatbg}{\widehat{\boldsymbol g}}
\newcommand{\hatbmu}{\widehat{\boldsymbol {\mu}}}

\newcommand{\vx}{ {\bf x} }
\newcommand{\vk}{ {\bf k} }
\newcommand{\vq}{ {\bf q} }
\newcommand{\vp}{ {\bf p} }

\newcommand{\vn}{ {\bf n} }

\newcommand{\lambdabar}{\overline{\lambda}}

\def\({\left(}
\def\){\right)}
\def\[{\left[}
\def\]{\right]}

\def\nn{\nonumber}
\def\bec{\begin{center}}
\def\eec{\end{center}}
\def\beq{\begin{equation}}
\def\eeq{\end{equation}}
\def\bea{\begin{eqnarray}}
\def\eea{\end{eqnarray}}
\newcommand{\cbar}{{\overline{c}}}
\newcommand{\myref}[1]{(\ref{#1})}

\title{Perturbative Renormalization of Lattice ${\cal N} = 4$
Super Yang-Mills Theory}

\author{Simon Catterall \\
Department of Physics, Syracuse University, Syracuse, NY 13244, USA}

\author{Eric Dzienkowski \\
Department of Physics, University of California Santa Barbara, CA 93106, USA}

\author{Joel Giedt \\
Department of Physics,  Applied Physics and Astronomy,  Rensselaer Polytechnic Institute, 110 8th Street, Troy NY 12065 USA}

\author{Anosh Joseph \\
Department of Physics, Syracuse University, Syracuse, NY 13244, USA}

\author{Robert Wells\\
DAMTP, CMS, University of Cambridge, Cambridge, CB3 0WA, UK}

\preprint{SU-4252-912}

\abstract{
We consider ${\cal N}=4$ super Yang-Mills theory on a four-dimensional lattice. The lattice formulation under consideration retains one exact supersymmetry at non-zero lattice spacing. We show that this feature combined with gauge invariance and the large point group symmetry of the lattice theory
ensures that the only counterterms that appear at any order in perturbation theory correspond to renormalizations of existing
terms in the bare lattice action. In particular we find that no mass terms are generated at any finite order of perturbation theory. We calculate
these renormalizations by examining the fermion and auxiliary boson
self energies at one loop and find that they all exhibit a common logarithmic
divergence which can be absorbed by a single wavefunction renormalization.
This finding implies that at one loop only a fine tuning of the
finite parts is required to regain full supersymmetry in the continuum limit.
}

\keywords{Lattice Quantum Field Theory, Supersymmetric Gauge Theory, Topological Field Theories, Extended Supersymmetry}

\begin{document}

%%%%%%%%%%%%%%%%%%%%%%%%%%%%%%%%%%%%%%%%%%%%%%%%%%%%%%%%%%%%%%%%%
\tableofcontents

%%%%%%%%%%%%%%%%%%%%%%%%%%%%%%%%%%%%%%%%%%%%%%%%%%%%%%%%%%%%%%%%%
\section{Introduction}
%%%%%%%%%%%%%%%%%%%%%%%%%%%%%%%%%%%%%%%%%%%%%%%%%%%%%%%%%%%%%%%%%

${\cal N} = 4$ supersymmetric Yang-Mills (SYM) theory in 
four dimensions is both a fascinating quantum field theory in its own right and in addition plays a crucial role in the well known AdS/CFT correspondence \cite{Maldacena:1997re, Itzhaki:1998dd} where it is thought to provide a dual description of type IIB string theory in $AdS_5 \times S_5$ space. 

The strong coupling, large-$N$ limit ($N$ being the number of colors) of this gauge theory has been extensively studied since the corresponding dual theory reduces to a weakly coupled supergravity theory, which describes the low energy limit of type IIB string theory. However it has proven difficult to go beyond this leading supergravity approximation in most situations. Having a lattice formulation of the super Yang-Mills theory would clearly be very advantageous as it would give a non-perturbative definition of the gauge theory and offer up new tools to investigate its strong coupling dynamics and thence the dynamics of the dual theory. Indeed such a lattice construction would allow for a systematic study of the classical and quantum string corrections to the supergravity solution. Unfortunately straightforward attempts to discretize the continuum theory are well known to break supersymmetry completely leading to a profusion of supersymmetry violating counterterms (four or six counterterms, depending on the gauge group, in the effective action whose couplings must be fine tuned to approach the correct continuum limit - see Ref.\cite{Elliott:2008jp}).

However, recent formulations of supersymmetric lattice theories, which retain exact supersymmetry at non zero lattice spacing offer the hope of at least partially evading these fine tuning problems - the existence of an exact lattice supersymmetry protecting the theory from many of these dangerous counterterms \cite{Catterall:2003wd, Sugino:2003yb, Cohen:2003xe, Cohen:2003qw, Catterall:2004np, D'Adda:2005zk, Kaplan:2005ta, Unsal:2006qp, Damgaard:2007be, Damgaard:2007xi, Catterall:2007kn}. See also the recent reviews \cite{Giedt:2009yd, Catterall:2009it} for further references. In addition,
there has recently 
been a great deal of other work, both theoretical and numerical,
focused on discrete formulations of $\cN=4$ SYM see Refs. \cite{Ishii:2008ib, Ishiki:2008te, Ishiki:2009sg, Nishimura:2009xm, Hanada:2010kt, Hanada:2010gs, Honda:2010nx}. 
These alternative approaches should be viewed as complementary to the
lattice construction described in this paper.

In the case of ${\cal N}=4$ SYM the corresponding lattice theory retains only one out of the sixteen continuum supersymmetric invariances and the question of how much fine tuning is required to take a continuum limit of this lattice theory targeting the usual ${\cal N}=4$ theory has been unclear up to this point in time. This paper aims to address this issue using both general arguments valid to all orders in perturbation theory and an explicit calculation of the renormalization of the lattice theory to one loop order.

We will argue quite generally that the symmetries of the lattice theory strongly constrain the possible counterterms that can arise as a result of quantum corrections; we find that the only relevant operators that can be induced via radiative effects correspond to renormalizations of four marginal operators already present in the tree level theory. These operators correspond to kinetic terms and we show, using a topological argument based on the exact (twisted) lattice supersymmetry, that no mass terms are induced to all orders in perturbation theory.

The remaining fine tuning question then hinges on what divergences
 can arise in the renormalization of these four bare couplings.
 We proceed to calculate these divergences at one loop using 
lattice perturbation theory. Exact lattice supersymmetry allows
 us to extract these leading divergences by examining  the renormalization
 of the three types of twisted fermion propagator and a single propagator
 for an auxiliary bosonic field. We show that all these exhibit a common
 logarithmic divergence at one loop.   The appearance of a 
single logarithmic divergence ensures that at one loop only
finite parts need to be fine tuned in order to
 regain full supersymmetry in the continuum limit. This is a huge
 advantage of this approach as compared to earlier efforts at
 constructing supersymmetric lattice theories in four dimensions.

We start with a discussion of the approach to supersymmetric lattices through discretization of a topologically twisted formulation of the super Yang-Mills theory, write down the lattice theory and discuss the constraints on its renormalization implied by the lattice symmetries. We then derive the Feynman rules governing the perturbative structure of the lattice theory and write down the diagrams needed to renormalize the theory at one loop. We compute the partition function at one loop and show that it is independent of any background fields and furthermore that this is true to all orders in perturbation theory. We then evaluate the one loop diagrams, extract their leading logarithmic divergences and compute the required renormalization of the lattice theory. We conclude with a summary of our main results.

%%%%%%%%%%%%%%%%%%%%%%%%%%%%%%%%%%%%%%%%%%%%%%%%%%%%%%%%%%%%%%%%%
\section{Twisted ${\cal N} = 4$, $d=4$ Super Yang Mills}
%%%%%%%%%%%%%%%%%%%%%%%%%%%%%%%%%%%%%%%%%%%%%%%%%%%%%%%%%%%%%%%%%

The key idea that allows us to construct a supersymmetric lattice theory that targets ${\cal N}=4$ super Yang-Mills in the (na\"ive) continuum limit is called topological twisting - see \cite{Catterall:2009it} and references therein\footnote{This approach to supersymmetric lattices has been shown to be entirely equivalent to the orbifold/deconstruction formulations pioneered by Kaplan, \"Unsal and collaborators \cite{Kaplan:2005ta} which predated the twisted constructions in the case of gauge theories. However for simplicity we will mostly be using the language of twisting in this paper.}. This twisting process, which in flat space can be thought of as merely a change of variables, exposes a scalar, nilpotent supercharge $\cQ$. It is the supersymmetry associated with this supercharge which can be implemented exactly in the lattice theory.

To understand how the twisted theory is constructed one needs to examine the relevant global symmetries of the continuum Yang-Mills theory. The Euclidean version of the $\cN = 4$ SYM theory on ${\mathbb R}^4$ can be obtained by dimensionally reducing $\cN = 1$ SYM theory on ${\mathbb R}^{10}$ down to ${\mathbb R}^4$. The ten-dimensional theory possesses an $SO(10)$ Euclidean (Lorentz) rotation group. After dimensional reduction it reduces to 
\bec
$SO(10)_E \rightarrow SO(4)_E \times SO(6)_I$~,
\eec
where $SO(4)_E \sim SU(2) \times SU(2)$ is the four-dimensional Euclidean (Lorentz) symmetry on ${\mathbb R}^4$ and $SO(6)_I \sim SU(4)_R$ is the global internal $R$-symmetry group of the dimensionally reduced theory. 

The basic idea of twisting is to decompose the fields of the theory in irreducible representations of a twisted rotation group which involves both the usual rotations and the R-symmetry\footnote{The ${\cal N}=4$ theory may be twisted in three inequivalent ways \cite{Yamron:1988qc, Vafa:1994tf, Marcus:1995mq} however only one of these - that due to Marcus and described in the text is ultimately compatible with discretization.}. The global $R$-symmetry group of the dimensionally reduced theory contains a subgroup $SO(4)_{R} \times U(1)$. To construct the relevant twist needed for the lattice theory, we adopt as twisted rotation group $SO(4)^\prime$ the diagonal subgroup of $SO(4)_E \times SO(4)_R$. The global symmetry is now given by
\bea
G' &=&  SO(4)' \times U(1) \nn \\
&\sim& SU(2)' \times SU(2)' \times U(1) \nn
\eea
Notice that since the $U(1)$ part of the global internal symmetry group is undisturbed, it remains as a global $R$-symmetry of the twisted theory.
The supercharges and fermions transform under the new rotation group as
\beq
SU(2)' \times SU(2)' \times U(1) \rightarrow (1,1)_{\frac{1}{2}} \oplus (2,2)_{-\frac{1}{2}} \oplus [(3,1) \oplus (1,3)]_{\frac{1}{2}} \oplus (2,2)_{-\frac{1}{2}} \oplus (1,1)_{\frac{1}{2}}~,
\eeq
or equivalently
\beq
SO(4)' \times U(1) \rightarrow 1_{\frac{1}{2}} \oplus 4_{-\frac{1}{2}} \oplus 6_{\frac{1}{2}} \oplus 4_{-\frac{1}{2}} \oplus 1_{\frac{1}{2}}~.
\eeq
As a result of this choice of embedding, the twisted theory contains supersymmetries and fermions in integer spin representations. They transform as scalars, vectors and higher rank $p$-form tensors:
\bec
  \begin{tabular}{ l  c  c  c  c  c  c  c  c r }
    $\textrm{Supercharges: }$ & $\cQ^{(0)}$ & $\oplus$ & $\cQ^{(1)}$ & $\oplus$ & $\cQ^{(2)}$ & $\oplus$ & $\cQ^{(3)}$ & $\oplus$ & $\cQ^{(4)}$\\ 
    $\textrm{Fermions: }$ & $\Psi^{(0)}$ & $\oplus$ & $\Psi^{(1)}$ & $\oplus$ & $\Psi^{(2)}$ & $\oplus$ & $\Psi^{(3)}$ & $\oplus$ & $\Psi^{(4)}$\\ 
  \end{tabular}
\eec
We parametrize the fermionic content of the theory by 
\beq
\Psi = \begin{cases}\eta &  1 \\ 
		    \psi_{\mu} &  4 \\
		    \chi_{\mu \nu} & 6 \\ 
		    \bar{\psi}_{\mu} \equiv \epsilon_{\mu \nu \rho \lambda}\xi_{\nu \rho \lambda} & 4 \\
		    \bar{\eta} \equiv \epsilon_{\mu \nu \rho \lambda} \psi_{\mu \nu \rho \lambda} & 1 \ \end{cases}
\eeq
The four gauge bosons transform as $(2, 2)$ under the twisted rotation group. We label them as a vector field $A_{\mu}$. Similarly, four of the six scalars of the theory are now elevated to the same footing as the gauge bosons; they also transform as $(2, 2)$ under the twisted rotation group. We label them as a vector field $B_{\mu}$. The two other scalars remain as singlets under the twisted rotation group. We label them by $\phi$ and $\bar{\phi}$. Thus the bosons of the twisted theory transform as:
\beq
SU(2)' \times SU(2)' \times U(1) \rightarrow (1, 1)_1 \oplus (2, 2)_0 \oplus (2, 2)_0 \oplus (1, 1)_{-1}~,
\eeq
or equivalently
\beq
SO(4)' \times U(1) \rightarrow 1_1 \oplus 4_0 \oplus 4_0 \oplus 1_{-1}~.
\eeq
We parametrize the bosonic content of the theory by 
\beq
\Phi = \begin{cases}\phi &  1 \\ 
		    A_{\mu} &  4 \\
		    B_{\mu} & 4 \\ 
		    \phib & 1 \ \end{cases}
\eeq
For future reference the fields of the twisted theory have the following mass dimensions:
\bec
  \begin{tabular}{ | l | c | c | c | c | c | c | c | c | c | r | }
    \hline
    $\textrm{Twisted field}$ & $A_{\mu}$ & $B_{\mu}$ & $\phi$ & $\bar{\phi}$ & $\chi_{\mu \nu}$ & $\psi_{\mu}$ & ${\bar{\psi}}_{\mu}$ & $\eta$ &  ${\bar{\eta}}$ \\ \hline
    $\textrm{Mass dimension}$ & 1 & 1 & 1 & 1 & 3/2 & 3/2 & 3/2 & 3/2 & 3/2 \\ \hline
  \end{tabular}
\eec
%The twist described above supports the existence of two Lorentz singlet supercharges transforming as $(1, 1)$ under the twisted Lorentz group. They are the spin-zero supercharge $Q^{(0)}$ associated with zero-form Grassmann $\Psi^{(0)}$ and the supercharge $^*Q^{(4)}$ associated with the Poincar\'e dual of the $4$-form Grassmann, $^{*}\Psi^{(4)}$. In this paper we choose to focus on the scalar nilpotent supersymmetry given by the linear combination
%\beq
%\cQ = Q^{(0)} + ^*Q^{(4)}~,
%\eeq
%where $^*Q^{(4)} = \frac{1}{4!} \epsilon_{\mu \nu \rho \lambda} Q_{\mu \nu \rho \lambda}$. The resulting scalar supercharge, $\cQ \sim (1, 1)$, is nilpotent, $\cQ^2 ~\cdot = 0$. It is also interesting to see that with this particular choice of twist, the resulting twisted fermions are just sufficient to saturate a single Dirac-K\"ahler field \cite{D'Adda:2005zk}.

In this paper we are interested in $Q^{(0)}$, the scalar supercharge which is nilpotent $(\cQ^{(0)})^2 ~\cdot = 0$. It is also interesting to see that with this particular choice of twist, the resulting twisted fermions are just sufficient to saturate a single Dirac-K\"ahler field \cite{D'Adda:2005zk}.

\section{Continuum Action and Nilpotent Supersymmetry}

Since the two vector fields $A_{\mu}$ and $B_{\mu}$ of the twisted theory transform the same way under the twisted rotation group we can describe the theory in a compact way if we combine the vector fields into a complex vector field $\cA_{\mu}$ \cite{Marcus:1995mq}\footnote{Throughout this paper we will be employing an hermitian basis for the generators satisfying $\Tr(T^aT^b)=\frac{1}{2}\delta^{ab}$.}:
\bea
\cA_{\mu} \equiv A_{\mu} + iB_{\mu}~,\\
\cAb_{\mu} \equiv A_{\mu} - iB_{\mu}~.
\eea
Using these connections
one can now define three covariant derivatives and field strengths:
\bea
D_{\mu} ~\cdot \equiv \partial_{\mu} + ig[A_{\mu}, ~\cdot~ ],~~~ F_{\mu \nu} \equiv -\frac{i}{g}[D_{\mu}, D_{\nu}]~,\\
\cD_{\mu} ~\cdot \equiv \partial_{\mu} + ig[\cA_{\mu}, ~\cdot~ ],~~~ \cF_{\mu \nu} \equiv -\frac{i}{g}[\cD_{\mu}, \cD_{\nu}]~,\\
\cDb_{\mu} ~\cdot \equiv \partial_{\mu} +ig[\cAb_{\mu}, ~\cdot~ ],~~~ \cFb_{\mu \nu} \equiv -\frac{i}{g}[\cDb_{\mu}, \cDb_{\nu}]~.
\eea
To make contact with the final lattice construction it is
useful to assemble
the complexified gauge fields and the two scalar fields into a single five-component complexified connection
\beq
\cA_{a} = \Big(\cA_{\mu} \equiv A_{\mu} + i B_{\mu},~~\cA_{5} \equiv A_5 + i B_5 \Big)~, ~~~a = 1, \cdots, 5\; ; \mu = 1, \cdots, 4
\eeq
where the fifth component $\cA_5 = \phi$ and $\cAb_5 = \phib$. Correspondingly the fermions can be packaged as five-dimensional scalar, vector and antisymmetric tensors $(\eta,\psi_a,\chi_{ab})$. The original twisted theory will then be obtained by simple dimensional reduction of a theory in five dimensions. A similar language arises in the orbifold construction of this theory where the fermions and bosons transform in representations of $SU(5)$.
In addition to these fields
we introduce one auxiliary bosonic scalar field $d$ for off-shell completion of the scalar supersymmetry.

The nilpotent scalar supersymmetry $\cQ$ (from now on we denote the scalar supersymmetry $\cQ^{(0)}$ by $\cQ$) now acts on these fields in a simple
manner
\bea
\cQ \cA_{a} &=& \psi_{a} \\
\cQ \psi_a &=& 0 \\
\cQ \cAb_a &=& 0 \\
\cQ \chi_{ab} &=& \cFb_{ab} \\
\cQ \eta &=& d \\
\cQ d &=& 0
\eea
The action of the twisted theory can now be expressed in a compact five-dimensional form, as a linear combination of $\cQ$-exact and $\cQ$-closed terms
\beq
S = \cQ \Lambda + S_{\cQ{\rm -closed}}~,
\eeq
where
\beq
\Lambda = \int\Tr \Big(\chi_{ab} \cF_{ab} - \frac{i}{g}\eta [\cDb_a, \cD_a] + \frac{1}{2} \eta d\Big)~,
\eeq 
and 
\beq
S_{\cQ{\rm -closed}} = - \frac{1}{2} \int ~\Tr \epsilon_{abcde} \chi_{de} \cDb_{c} \chi_{ab}~.
\eeq
The invariance of the $\cQ$-closed term is a result of the Bianchi identity (or Jacobi identity for covariant derivatives) 
\beq
\epsilon_{abcde}\cDb_{c}\cFb_{de} = -\frac{i}{g}\epsilon_{abcde}{[} \cDb_c, {[} \cDb_d, \cDb_e]] = 0~.
\eeq
Carrying out the $\cQ$-variation and subsequently eliminating the auxiliary field $d$ using the equation of motion
we can write down the action in terms of the propagating fields.
\bea
S &=& \int\Tr \Big( \cFb_{ab} \cF_{ab} + \frac{1}{2g^2} [\cDb_a, \cD_a]^2 - \chi_{ab} \cD_{[a}\psi_{b]} - \eta\cDb_a \psi_a \nn \\
&&~~~~~~~~~~~- \frac{1}{2}\epsilon_{abcde} \chi_{de} \cDb_{c} \chi_{ab}\Big).~~~
\eea
After a redefinition of the fields $g\eta \rightarrow \eta$, $g\psi_a \rightarrow \psi_a$, $g\chi_{ab} \rightarrow \chi_{ab}$ and $g\cA_a \rightarrow \cA_a$ we have
\bea
S &=& \frac{1}{g^2}\int\Tr \Big( -[\cDb_a, \cDb_b][\cD_a, \cD_b] + \frac{1}{2} [\cDb_a, \cD_a]^2 - \chi_{ab} \cD_{[a}\psi_{b]} - \eta\cDb_a \psi_a \nn \\
&&~~~~~~~~~~~- \frac{1}{2}\epsilon_{abcde} \chi_{de} \cDb_{c} \chi_{ab}\Big).~~~
\label{eq:compact-five-d-action}
\eea
The target twisted theory in four dimensions can be obtained by dimensional reduction of this theory along  the 5th direction. We write down the decomposition of five-dimensional fields into four-dimensional fields as follows
\bea
\cA_a &\rightarrow& \cA_{\mu} \oplus \phi \\
\cF_{ab} &\rightarrow& \cF_{\mu \nu} \oplus \cD_{\mu} \phi \\
{[}\cDb_a, \cD_a ] &\rightarrow& {[}\cDb_{\mu}, \cD_{\mu}] \oplus -[\overline{\phi}, \phi] \\
\psi_a &\rightarrow& \psi_{\mu} \oplus \overline{\eta} \\
\chi_{ab} &\rightarrow& \chi_{\mu \nu} \oplus \overline{\psi}_{\mu}
\eea
The action (\ref{eq:compact-five-d-action}), after dimensional reduction, yields
\bea
S &=& \frac{1}{g^2}\int \Tr \Big(-[\cDb_{\mu}, \cDb_{\nu}][\cD_{\mu}, \cD_{\nu}] + \frac{1}{2} [\cDb_{\mu}, \cD_{\mu}]^2 + \frac{1}{2}{[}\phib, \phi]^2 - (\cDb_{\mu}\phib) (\cD_{\mu}\phi) - \chi_{\mu \nu} \cD_{[\mu} \psi_{\nu]} \nn \\
\label{eq:susy-lattice-twist-action}
&& -\psib_{\mu} \cD_{\mu} \etab - i\psib [\phi, \psi_{\mu}] - \eta \cDb_{\mu} \psi_{\mu} - i\eta {[}\phib, \etab] - \chi^*_{\mu \nu} \cDb_{\mu} \psib_{\nu} - \frac{i}{2}\chi^*_{\mu \nu} {[}\phib, \chi_{\mu \nu}]\Big),
\eea
where the last two terms arise from the dimensional reduction of the $\cQ$-closed term with $\chi^*$, the Hodge dual of $\chi$,
defined as $\chi^*_{\mu \nu} = \frac{1}{2} \epsilon_{\mu \nu \rho \lambda} \chi_{\rho \lambda}$ and $\psib_\mu=\frac{1}{2}\chi_{5\mu}$.  

This action can be identified with the twisted $\cN=4$ SYM action in four dimensions written down by Marcus \cite{Marcus:1995mq}, up to a trivial rescaling of the fields (with a gauge parameter $\alpha = 1$ in \cite{Marcus:1995mq}). It is important to note that in flat space, this twisted action is just a rewriting of the usual $\cN=4$ SYM theory in four dimensions and is physically equivalent to it.

%%%%%%%%%%%%%%%%%%%%%%%%%%%%%%%%%%%%%%%%%%%%%%%%%%%%%%%%%%%%%%%%%
\section{Lattice Theory}
%%%%%%%%%%%%%%%%%%%%%%%%%%%%%%%%%%%%%%%%%%%%%%%%%%%%%%%%%%%%%%%%%

Discretization of the twisted theory described in the previous section proceeds straightforwardly; complex continuum gauge fields are represented as complexified Wilson gauge links $\cU_a(\vn)$ living on links $\mu_a,\,a=1\ldots 5$ of a four-dimensional lattice. Since there are five such vectors it should be clear that this lattice will have five basis vectors. To ensure that the lattice theory enjoys a maximal symmetry we would like these basis vectors to all be equivalent. This requirement means that the lattice must possess an $S^5$ point group symmetry (the Weyl group of $SU(5)$). The unique solution to these constraints in four dimensions is the so-called $A_4^*$ lattice. This will hence be the underlying lattice used in our work.

A specific basis for the $A_4^*$ lattice is given in the form of five lattice vectors
\bea
\hatbe_1 &=&  \Big(\frac{1}{\sqrt{2}}, \frac{1}{\sqrt{6}}, \frac{1}{\sqrt{12}}, \frac{1}{\sqrt{20}}\Big)\\
\hatbe_2 &=& \Big(-\frac{1}{\sqrt{2}}, \frac{1}{\sqrt{6}}, \frac{1}{\sqrt{12}}, \frac{1}{\sqrt{20}}\Big)\\
\hatbe_3 &=& \Big(0, -\frac{2}{\sqrt{6}}, \frac{1}{\sqrt{12}}, \frac{1}{\sqrt{20}}\Big)\\
\hatbe_4 &=& \Big(0, 0, -\frac{3}{\sqrt{12}}, \frac{1}{\sqrt{20}}\Big)\\
\hatbe_5 &=& \Big(0, 0, 0, -\frac{4}{\sqrt{20}}\Big).
\eea
These lattice vectors connect the center of a $4$-simplex to its five corners. They are related to the $SU(5)$ weights of the $5$ representation. 
The unit cell of the $A_4^*$ lattice is a compound of two $4$-simplices corresponding to the $5$ (formed by the basis vectors $\hatbe_m$) and $\overline{5}$ (formed by the basis vectors $-\hatbe_m$) representations of $SU(5)$.
The basis vectors satisfy the relations 
\beq
\sum_{m=1}^{5} \hatbe_m = 0;~~\hatbe_m \cdot \hatbe_n = \Big(\delta_{mn} - \frac{1}{5}\Big);~~\sum_{m=1}^{5}(\hatbe_m)_{\mu}(\hatbe_m)_{\nu} = \delta_{\mu \nu}; ~~\mu, \nu = 1, \cdots, 4.
\eeq
Notice also that $S^5$ is a subgroup of the twisted rotation symmetry group $SO(4)^\prime$ and that the lattice fields transform in reducible representations of this discrete group - for example the vector $\cA_a$ decomposes into a four component vector $\cA_\mu$ and a scalar field $\phi$ under $SO(4)^\prime$. Invariance of the lattice theory with respect to these discrete rotations then guarantees that the theory will inherit full invariance under twisted rotations in the continuum limit.

Proceeding in this manner it is possible to assign all the remaining fields to links on the $A_4^*$ lattice. Since $\psi_a(\vn)$ is a superpartner of $\cU_a(\vn)$ it must also reside on the link connecting $\vn \to \vn+\hatbe_a$. Conversely the field $\cU^{\dagger}_a(\vn)$ resides on the oppositely oriented link from $\vn \to \vn-\hatbe_a$. The ten fermions $\chi_{ab}(\vn)$ are then chosen to
reside on new fermionic links $\vn+\hatbe_m+\hatbe_n\to \vn$ while the singlet fermionic field $\eta(\vn)$ is assigned to the degenerate link consisting of a single site $\vn$.

The action of the theory takes the following form
\bea
\label{eq:action-on-a4-star}
S &=& \frac{1}{g^2}\sum_{\vn, a,b,c,d,e} ~\Big\{ \cQ ~\Tr \Big[-i\chi_{ab} \cD^{(+)}_a \cU_b(\vn) - \eta(\vn) \Big(i\cD^{\dagger(-)}_a \cU_a(\vn) - \frac{1}{2}d(\vn) \Big)\Big]\nn \\
&& - \frac{1}{2} \Tr \epsilon_{abcde} \chi_{de}(\vn + \hatbmu_a + \hatbmu_b + \hatbmu_c) \cD^{\dagger(-)}_{c} \chi_{ab}(\vn + \hatbmu_c)\Big\}~.
\eea
where the lattice field strength is given by
\beq
\cF_{ab}(\vn) \equiv -\frac{i}{g}\cD^{(+)}_a \cU_b(\vn) = -\frac{i}{g}\Big(\cU_a(\vn) \cU_b(\vn + \hatbmu_a) - \cU_b(\vn) \cU_a(\vn + \hatbmu_b)\Big).
\eeq
and the covariant difference operators appearing in this expression
are given by
\bea
\cD_c^{(+)} f(\vn) &=& \cU_c(\vn) f(\vn + \hatbmu_c) - f(\vn) \cU_c(\vn), \\
\cD_c^{(+)} f_d(\vn) &=& \cU_c(\vn) f_d(\vn + \hatbmu_c) - f_d(\vn) \cU_c(\vn + \hatbmu_d), \\
\cD_c^{\dagger(-)} f_c(\vn) &=& f_c(\vn)\cU^{\dagger}_c(\vn) - \cU^{\dagger}_c(\vn - \hatbmu_c) f_c(\vn - \hatbmu_c), \\
\cD_c^{\dagger(-)} f_{ab} (\vn) &=& f_{ab}(\vn) \cU^{\dagger}_c(\vn - \hatbmu_c) - \cU^{\dagger}(\vn + \hatbmu_a + \hatbmu_b - \hatbmu_c) f_{ab}(\vn - \hatbmu_c).
\eea
Notice that these definitions reduce to the usual adjoint covariant derivative in the na\"ive continuum limit corresponding to
$\cU_a=I+\cA_a(x)+\ldots$ 
and furthermore guarantee that the resultant discrete expressions transform covariantly under lattice gauge transformation. 
Furthermore, this use of forward and backward difference operators guarantees that the solutions of the theory map one-to-one with the solutions of the continuum theory and hence fermion doubling problems are evaded \cite{Rabin:1981qj}. Indeed, by introducing a lattice with half the lattice spacing one can map this Dirac-K\"{a}hler fermion action into the action for staggered fermions \cite{Banks:1982iq}. 

It is important to realize
that the vectors $\vn$, $\vn+\hatbmu_a$ etc appearing in this action do not correspond to the positions in spacetime of sites and links of the original $A_4^*$ lattice itself -- instead they span an abstract hypercubic lattice whose sites and links are given by integer valued lattice vectors. (Which are related to the $\br$-charges defined in the orbifold formulation \cite{Kaplan:2005ta}.) These 4-vectors $\hatbmu_a$ are defined as
\bea
\label{eq:mu-vectores}
\hatbmu_1 &=& (1, 0, 0, 0)\nn \\
\hatbmu_2 &=& (0, 1, 0, 0)\nn \\
\hatbmu_3 &=& (0, 0, 1, 0) \\
\hatbmu_4 &=& (0, 0, 0, 1)\nn \\
\hatbmu_5 &=& (-1, -1, -1, -1)\nn
\eea
The integer-valued lattice site $\vn$ can be related to the physical location in spacetime using the $A_4^*$ basis vectors $\hatbe_a$. 
\beq
\bR = a \sum_{\nu =1}^4 (\mu_{\nu} \cdot \vn)\hatbe_{\nu} = a \sum_{\nu =1}^{4}n_{\nu}\hatbe_{\nu}~,
\eeq
where $a$ is the lattice spacing. On using the fact that $\sum_{m}\hatbe_m = 0$, we can show that a small lattice displacement of the form $d\vn = \hatbmu_m$ corresponds to a spacetime translation by $(a\hatbe_m)$:
\beq
d\bR = a \sum_{\nu =1}^{4}(\mu_{\nu} \cdot d \vn)\hatbe_{\nu} = a \sum_{\nu =1}^{4} (\hatbmu_{\nu} \cdot \hatbmu_m)\hatbe_{\nu} = a \hatbe_m~.
\eeq

The supersymmetry transformations on the lattice fields are almost identical to their continuum counterparts:
\bea
\cQ \cU_a(\vn) &=& \psi_{a}(\vn) \\
\cQ \psi_a(\vn)  &=& 0 \\
\cQ \cU^\dagger_a(\vn)  &=& 0 \\
\cQ \chi_{ab}(\vn)  &=& i(\cD_a^{(+)} \cU_b)^{\dagger}(\vn)  \\
\cQ \eta (\vn) &=& d \\
\cQ d (\vn) &=& 0
\eea
After the $\cQ$-variation, as performed in the continuum, and integrating out the auxiliary field $d$, the final lattice action is
\bea
\label{eq:final-action}
S &=& \frac{1}{g^2}\sum_{\vn,a,b,c,d,e} \Tr \Big[\Big(\cD_a^{(+)}\cU_b(\vn)\Big)^{\dagger}\Big(\cD^{(+)}_a \cU_b(\vn)\Big) + \frac{1}{2} \Big(\cD^{\dagger(-)}_{a} \cU_{a}(\vn)\Big)^2 - \chi_{ab}(\vn) \cD^{(+)}_{[a} \psi_{b]}(\vn) \nn \\
&&- \eta(\vn) \cD^{\dagger(-)}_{a} \psi_{a}(\vn) - \frac{1}{2} \epsilon_{abcde} \chi_{de}(\vn + \hatbmu_a + \hatbmu_b + \hatbmu_c) \cD^{\dagger(-)}_{c} \chi_{ab}(\vn + \hatbmu_c)\Big].
\eea
To see that this action targets the continuum twisted theory one needs to expand $\cU_a$ about the unit matrix \cite{Kaplan:2005ta}\footnote{To leading order this is equivalent to the more conventional expression $\cU_a(x)=\frac{1}{a}e^{ia\cA_a(x)}$. We will see that the 
linear representation offers important advantages over the exponential
in our later calculations.}
\bea
\label{eq:u}
\cU_a (\vn) &=& \frac{1}{a}{\mathbb I}_N + i\cA_a(\vn)~, \\
\label{eq:cu}
\cU^{\dagger}_a (\vn) &=& \frac{1}{a}{\mathbb I}_N - i\cAb_a(\vn)~.
\eea
While the supersymmetric invariance of the $\cQ$-exact term is manifest in the lattice theory it is not immediately clear that the $\cQ$-closed term remains supersymmetric after discretization. Remarkably, this can be shown using a remarkable property of the discrete field strength which can be shown to satisfy an exact Bianchi identity just as for the continuum \cite{Aratyn:1984bd}.
\beq
\epsilon_{abcde}\cD^{\dagger(-)}_c\cF^{\dagger}_{ab}(\vn + \hatbmu_c)=0
\eeq

%%%%%%%%%%%%%%%%%%%%%%%%%%%%%%%%%%%%%%%%%%%%%%%%%%%%%%%%%%%%%%%%%
\section{Renormalization - General Analysis}
\label{sec:general}
%%%%%%%%%%%%%%%%%%%%%%%%%%%%%%%%%%%%%%%%%%%%%%%%%%%%%%%%%%%%%%%%%

Power counting reveals that the continuum four-dimensional theory has an infinite number of superficially divergent Feynman diagrams occurring at all orders of perturbation theory. Of course in the continuum target  theory all of these potential divergences cancel between diagrams to render the quantum theory finite. However, since the lattice theory does not possess all the supersymmetries of the continuum theory, it is not clear how many of these will continue to cancel in the lattice theory.

Before we embark on a general perturbative analysis of this lattice theory it is instructive to try to ascertain what kinds of counter terms are permitted by the lattice symmetries. In the case of $A_4^{*}$ lattice, these symmetries are
\begin{itemize}
\item[{\bf a)}] Exact  $\cQ$ supersymmetry. 
\item [{\bf b)}] Gauge invariance
\item [{\bf c)}] $S_5$ point group symmetry and discrete translations.
\end{itemize}
In fact, other than exact lattice supersymmetry, the $U(N)$ lattice gauge theory also has a second fermionic  symmetry,  given by
\beq
\eta(\vn) \rightarrow \eta (\vn) + \epsilon {\mathbb I}_N, \qquad  \delta (\rm all \; other \; fields) =0 
\eeq
where $\epsilon$ is an infinitesimal Grassmann parameter. Thus, we extend our list to include 
\begin{itemize}
\item[{\bf d)}] Fermionic shift symmetry
\end{itemize}

In practice we are primarily interested in  relevant or marginal  operators; that is operators whose mass dimension is less than or equal to four. We will see that the set of relevant counterterms in the lattice theory is rather short -- the lattice symmetries, gauge invariance in particular, being extremely restrictive in comparison to the equivalent situation in the continuum. The argument starts by assigning canonical dimensions to the fields 
$[\cU_a]=1$, $[\Psi]=\frac{3}{2}$ and $[\cQ]=\frac{1}{2}$ where $\Psi$ stands for any of the twisted fermion fields $(\lambda,\psi_a,\chi_{ab})$. Invariance under $\cQ$ restricts the possible counterterms to be either of a  $\cQ$-exact form, or  of  $\cQ$-closed form. There is only one $\cQ$-closed operator permitted by the lattice symmetries and it is already present in our bare lattice action. A possible renormalization of this fermion kinetic term is hence allowed. Beyond that the exact lattice supersymmetry forces us to look at the set of $\cQ$-exact counterterms.    

Any such counterterm must be of the form $\mathcal{O}=\cQ\Tr(\Psi f(\cU,\cU^{\dagger}))$. There are thus no terms permitted by symmetries with dimension less than two. In addition gauge invariance tells us that each term must correspond to the trace of a closed loop on the lattice. The smallest dimension gauge invariant operator is then just $\cQ(\Tr\psi_a\cU^{\dagger}_a)$. But this vanishes identically since both $\cU^{\dagger}_a$ and $\psi_a$ are singlets under $\cQ$. No dimension $\frac{7}{2}$ operators can be constructed with this structure and we are left with just dimension four counterterms. Notice, in particular that lattice symmetries permit no simple fermion bi-linear mass terms.  However, gauge invariant fermion bi-linears with link field insertions are possible and their effect should be accounted for carefully.  

Possible  dimension four operators are, schematically, 
\begin{eqnarray}
L_1&=& g^{-2}\cQ \Tr (\chi_{ab}\cU_a \cU_b)  \cr
L_2&=&  g^{-2}\cQ \Tr (\eta \cD^{\dagger}_a \cU_a )\cr 
L_3&=&  g^{-2}\cQ \Tr (\eta \cU_a \cU^{\dagger}_a)\cr
L_4&=&  g^{-2}\cQ \Tr (\eta)\Tr(\cU_a \cU^{\dagger}_a)
\label{ops}
\end{eqnarray}
The  first operator  can be simplified on account of the antisymmetry of $\chi_{ab}$ to simply $\cQ(\chi_{ab}\cF_{ab})$, which again is nothing but 
one of the continuum $\cQ$-exact terms present in the bare action. The second term  in (\ref{ops}) also corresponds to one of the $\cQ$-exact terms in the bare action. However the third term $L_3$  is a new operator not present in the bare Lagrangian and the same is true for the final double-trace operator $L_4$. Both of these operators transform non-trivially under the fermionic shift symmetry, but a linear combination  of the two 
\beq 
D =  L_3 - \frac{1}{N} L_4  
\label{ren4}
\eeq
is invariant under the shift symmetry with $N$ the rank of the gauge group  $U(N)$. 
 
By these arguments it appears that the only relevant counterterms correspond to renormalizations of operators already present in the bare action  together with $D$. This is quite remarkable. The most general form for the renormalized lattice Lagrangian is hence 
\bea
\label{eq:general-form}
{\cal L} &=& \sum_{\vn,a,b,c,d,e} ~\Big\{ \cQ ~\Tr \Big[-i\alpha_1 \chi_{ab} \cD^{(+)}_a \cU_b(\vn) -i \alpha_2 \eta(\vn) \cD^{\dagger(-)}_a \cU_a(\vn) +\frac{\alpha_3}{2}\eta(\vn)d(\vn)\Big]\nn \\
&& - \frac{\alpha_4}{2} \Tr \epsilon_{abcde} \chi_{de}(\vn + \hatbmu_a + \hatbmu_b + \hatbmu_c) \cD^{\dagger(-)}_{c} \chi_{ab}(\vn + \hatbmu_c)\Big\}~ + \cQ \beta D~,
\eea
where $(\alpha_i,i=1\ldots 4)$ and $\beta$ are dimensionless numbers taking values $(1,1,1,1)$ and $0$ respectively in the classical lattice theory. 
Thus it 
appears that at most four dimensionless
ratios of these couplings might need to be tuned to approach $\cN=4$ Yang-Mills in the continuum limit. Furthermore,
since these operators are dimension four we expect this tuning to be at worst logarithmic in the cut-off.

In order to see the explicit form of the  $D$ operator close to the continuum limit, we expand the action around  $\cU_m(\vn)= \frac{1}{a}{\mathbb I}$. The result is 
\beq 
L_4 \sim \frac{1}{a}\left[ \Tr \eta(\vn)  (\sum_{m=1}^5 \psi^m(\vn) )   - \frac{1}{N} \Tr \eta(\vn) \Tr (\sum_{m=1}^5 \psi^m(\vn) )  + \right] 
\ldots
\eeq
where ellipsis are dictated by supersymmetry. The reader will immediately realize that $ (\sum_{a=1}^5 \psi_a)$ is nothing but the $S_5$  (and twisted $SO(4)'$) singlet contained in the reducible representation $\psi_a$. Indeed, it should be clear that it is the only field that could form a fermion mass term by pairing  with $\eta$. 

This is about as far as we can go by just using the lattice symmetries. We now turn to a full perturbative analysis to determine how the couplings $(\alpha_i, \beta)$ evolve with cut-off.

%%%%%%%%%%%%%%%%%%%%%%%%%%%%%%%%%%%%%%%%%%%%%%%%%%%%%%%%%%%%%%%%%
\section{Propagators and Vertices}
%%%%%%%%%%%%%%%%%%%%%%%%%%%%%%%%%%%%%%%%%%%%%%%%%%%%%%%%%%%%%%%%%

In this section we derive the propagators and vertices of the gauge-fixed $\cN =4$, $d=4$ SYM theory on $A_4^*$ lattice. Then we write down the one loop diagrams relevant for the renormalization of the theory.

Upon rewriting the field strength and covariant derivatives in terms of the bosonic link fields $\cU_a(\vn)$, the classical lattice
action (\ref{eq:final-action}) takes the form
\beq
S = S_B + S_F + S_c~,
\eeq
where
\bea
S_B &=& \frac{1}{g^2}\sum_{\vn,a,b} \Tr \Big[\Big(\cD_a^{(+)}\cU_b(\vn)\Big)^{\dagger}\Big(\cD^{(+)}_a \cU_b(\vn)\Big) + \frac{1}{2} \Big(\cD^{\dagger(-)}_{a} \cU_{a}(\vn)\Big)^2\Big]\nn \\
&=& \frac{1}{g^2}\sum_{\vn,a,b} \Tr \Big[\Big(\cU^{\dagger}_b(\vn + \hatbmu_a)\cU^{\dagger}_a(\vn) - \cU^{\dagger}_a(\vn + \hatbmu_b)\cU^{\dagger}_b(\vn)\Big)\Big(\cU_{a}(\vn)\cU_{b}(\vn + \hatbmu_a) - \cU_b(\vn)\cU_a(\vn + \hatbmu_b)\Big)\nn \\
&&+ \frac{1}{2} \Big(\cU_a(\vn)\cU^{\dagger}_a(\vn) - \cU^{\dagger}_a(\vn - \hatbmu_a)\cU_a(\vn - \hatbmu_a)\Big)^2\Big]~,\nn
\eea
\bea
\label{eq:final-action-U}
S_F &=& -\frac{1}{g^2}\sum_{\vn,a,b,c,d} \Tr \frac{1}{2}(\delta_{ac}\delta_{bd} - \delta_{ad}\delta_{bc}) \Big[\chi_{ab}(\vn)\Big(\cU_c(\vn)\psi_d(\vn + \hatbmu_c) - \psi_d(\vn)\cU_c(\vn + \hatbmu_d)\Big)\Big] \nn \\
&& + \eta(\vn)\Big(\psi_a(\vn)\cU^{\dagger}_a(\vn) - \cU^{\dagger}_a(\vn - \hatbmu_a)\psi_a(\vn - \hatbmu_a)\Big)~,
\eea
and
\bea
S_c &=& -\frac{1}{2g^2}\sum_{\vn,a,b,c,d,e} \Tr \epsilon_{abcde}\Big(\chi_{de}(\vn + \hatbmu_a + \hatbmu_b + \hatbmu_c)\nn \\
&&~~~~~~\times \Big[\chi_{ab}(\vn + \hatbmu_c)\cU^{\dagger}_c(\vn) - \cU^{\dagger}_c(\vn + \hatbmu_a + \hatbmu_b)\chi_{ab}(\vn)\Big]\Big)~.
\eea
To proceed further we expand the $\cU_a$ fields around unity as in eqn.~\ref{eq:u}. Notice that this expansion point is but one of an infinite number of classical vacuum solutions -- the full moduli space of the lattice theory corresponds to the set of all bosonic field variables $\cU_a(\vn)$ such that 
\bea
\label{eq:moduli-space-lattice}
0 &=& \sum_{\vn,a,b} \Tr \Big[\Big(\cU^{\dagger}_b(\vn + \hatbmu_a)\cU^{\dagger}_a(\vn) - \cU^{\dagger}_a(\vn + \hatbmu_b)\cU^{\dagger}_b(\vn)\Big)\Big(\cU_{a}(\vn)\cU_{b}(\vn + \hatbmu_a) - \cU_b(\vn)\cU_a(\vn + \hatbmu_b)\Big)\nn \\
&&+ \frac{1}{2} \Big(\cU_a(\vn)\cU^{\dagger}_a(\vn) - \cU^{\dagger}_a(\vn - \hatbmu_a)\cU_a(\vn - \hatbmu_a)\Big)^2\Big]~.\nn
\eea
These equations possess a large class of solutions corresponding to constant diagonal matrices modulo gauge transformations. We will use this additional freedom later when we compute the one loop contribution to the effective action of the theory.

%%%%%%%%%%%%%%%%%%%%%%%%%%%%%%%%%%%%%%%%%%%%%%%%%%%%%%%%%%%%%%%%%
\subsection{The Bosonic Propagators}
%%%%%%%%%%%%%%%%%%%%%%%%%%%%%%%%%%%%%%%%%%%%%%%%%%%%%%%%%%%%%%%%%

As usual it is easiest to compute the Feynman diagrams in momentum space. On the $A_4^*$ lattice a generic field $\Phi(\vx)$ has Fourier expansion
\beq
\Phi(\vx) = \frac{1}{(La)^4} \sum_{\vp}e^{i \vp \cdot \vx} \Phi_{\vp}
\label{ft}
\eeq
where $\vx=a\sum_{a=1}^4 n_a \hatbe_a$ denotes the position on $A_4^*$ and the momenta lie on the dual lattice given by $\vp=\frac{2\pi}{La}\sum_{a=1}^4 m_a \hatbg_a$ (for a lattice with spacing $a$ and length $L$). The dual basis vectors $\hatbg_a,a=1\ldots 4$ satisfy 
\beq
\hatbe_a .\hatbg_b=\delta_{ab}
\eeq
On an $L^4$ lattice both sets of lattice coordinates $n_a$, $m_a$  take integer values in the range $-L/2+1,\ldots,L/2$. We will assume periodic boundary conditions in all directions in this paper. Eqn.~\ref{ft} implies that fields are automatically invariant under translations by a lattice
length in any direction and a field shifted by one of the basis vectors can be expressed as\footnote{For simplicity we will adopt the convention
that momentum sums $\sum_k$ automatically include the $1/(La)^4$ 
normalization factor.}
\beq
\Phi(\vx+\hatbe_a)=\sum_{\vp}e^{ip_a}e^{i \vp \cdot \vx} \Phi_{\vp}
\eeq
where $p_a=\frac{2\pi}{L}m_a$. The only remaining is the question of how to deal with shifts in the lattice action associated with the additional $\hatbe_5$ vector. However, the solution is simple: since $\sum_{a=1}^5\hatbe_a=0$ we simply replace any $\hatbe_5$ shift encountered in the action by the equivalent shift $-\sum_{a=1}^4 \hatbe_a$. One might have worried about an apparent lack of rotational invariance associated with the na\"ive continuum limit of terms in the action which resemble $\sum_{a=1}^5 \sin^2{p_a}$ However, putting $p_a=\vp.\hatbe_a$ and taking the na\"ive continuum limit this becomes 
\beq
\sum_{a=1}^5 p_a^2=
\sum_{\mu,\nu}^4\sum_{a=1}^5 p_\mu p_\nu \hatbe^\mu_a \hatbe^\nu_a=
\sum_{\mu}^4 p_\mu^2
\eeq
which has the correct rotationally invariant form since the Greek indices refer to a Cartesian basis.
 
Using these ideas the bosonic action when expanded around (\ref{eq:u}) and (\ref{eq:cu}) gives the following second-order term in Fourier space
\bea
S^{(2)}_{B} &\approx& 2\sum_{\vk,a,b} \Tr \Big(\cAb_a(\vk)\Big[\delta_{ab}f_c(\vk)f^*_c(\vk) - f^*_a(\vk)f_b(\vk) \Big]\cA_b(-\vk) \nn \\
&&+ B_a(\vk)~\Big[f^*_a(\vk)f_b(\vk)\Big]~B_b(-\vk)\Big)~,
\eea
where
\beq
f_a(\vk) = (e^{i k_a}-1).
\eeq
We need to gauge-fix the bosonic action before we derive the propagators. A natural gauge-fixing choice would be an obvious generalization of Lorentz gauge-fixing \cite{Marcus:1995mq}
\beq
G(\vn) = \sum_{a} \Big(\partial_a^{(-)}\cA_a(\vn) + \partial_a^{(-)}\cAb_a(\vn)\Big)~.
\label{gfixf}
\eeq
This gauge-fixing choice adds the following term to the bosonic action at quadratic order
\beq
S_{GF}= \frac{1}{4 \alpha} \sum_{\vn} G^2(\vn)
= \frac{1}{\alpha} \sum_{\vn, a} \Tr (\partial^{(-)}_a A_a(\vn))^2,
\eeq
where $\partial^{(-)}_a f(\vn) = f(\vn) - f(\vn - \hatbmu_a)$. On using the relation $\sum_{\vn}(\partial^{(+)}_a f(\vn))g(\vn) = -\sum_{\vn}f(\vn)\partial^{(-)}_a g(\vn)$, the gauge-fixing term becomes
\beq
S_{GF} = -\frac{1}{\alpha} \sum_{\vn, a,b} \Tr A_a(\vn)\partial^{(+)}_{a}\partial^{(-)}_bA_b(\vn).
\eeq
In momentum space it becomes
\bea
S_{GF} &=&  \frac{1}{\alpha} \sum_{\vk, a, b} \Tr A_a(\vk) f^*_a(\vk)f_b(\vk) A_b(-\vk).
\eea
Thus the gauge-fixed bosonic action to quadratic order is
\bea
S^{(2)}_{B} + S_{GF}&\approx& 2 \sum_{\vk, a, b, c} \Tr
\Big(A_a(\vk)~\Big[\delta_{ab} f_c(\vk)f^*_c(\vk) - \Big(1 -
  \frac{1}{2\alpha}\Big) f^*_a(\vk)f_b(\vk)\Big]~A_b(-\vk) \nn \\
&&+ B_a(\vk)~\Big[\delta_{ab} f_c(\vk)f^*_c(\vk)\Big]~B_b(-\vk)\Big).
\eea
The choice $\alpha = 1/2$ makes the above expression diagonal
\bea
S^{(2)}_{B} &\approx& 2 \sum_{\vk, a, b, c} \Tr \cAb_a(\vk)~[\delta_{ab} f_c(\vk)f^*_c(\vk)]~\cA_b(-\vk)\nn \\
&=& 2 \sum_{\vk, a, b} \Tr \Big[\cAb_a(\vk)\delta_{ab} \Big( 4 \sum_c \sin^2\Big(\frac{k_c}{2}\Big)\Big)\cA_b(-\vk)\Big]~.
\eea
Putting in the trace (using the convention $\Tr (T^A T^B) = \frac{1}{2}\delta_{AB}$) the quadratic bosonic action can be written as
\beq
S^{(2)}_{B} \approx \sum_{\vk, a, b}\cAb^A_a(\vk)M^{AB}_{ab}(\vk)\cA^B_b(-\vk)~,
\eeq
where $M^{AB}_{ab}(\vk) = \widehat{\vk}^2 \delta_{ab}\delta_{AB}$, with $\widehat{\vk}^2 = 4\sum_c \sin^2\Big(\frac{k_c}{2}\Big)$.
Thus only the $\cA \cAb$ propagator is non-zero and it is given by (See figure~\ref{bosonic}.)
\beq
\langle \cA_a^{A}(-\vk) \cAb_b^B(\vk)\rangle = \delta_{ab}\delta_{AB} \frac{1}{\widehat{\vk}^2}~.
\eeq 
\FIGURE[t]{
\includegraphics[scale=1.0]{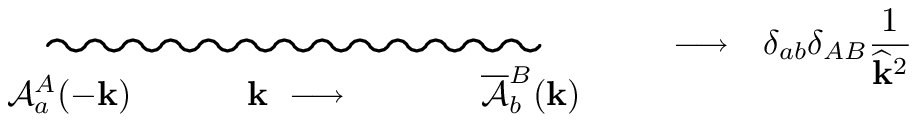}
\caption{\label{bosonic}The bosonic propagator.}}

%%%%%%%%%%%%%%%%%%%%%%%%%%%%%%%%%%%%%%%%%%%%%%%%%%%%%%%%%%%%%%%%%
\subsection{The Fermionic Propagators}
%%%%%%%%%%%%%%%%%%%%%%%%%%%%%%%%%%%%%%%%%%%%%%%%%%%%%%%%%%%%%%%%%

The fermionic part of the action is of the form
\beq
S_F = -\frac{1}{g^2}\sum_{abcde}\Big( \chi_{ab} \cD^{(+)}_{[a}\psi_{b]} +  \eta \cD_a^{\dagger(-)}\psi_a + \frac{1}{2} \epsilon_{abcde}\chi_{ab}\cD_c^{\dagger(-)}\chi_{de}\Big)
\eeq
Explicitly, we have
\bea
S_F &=& -\frac{1}{g^2}\sum_{\vn, a, b, c, d, e} \Tr \Big( \chi_{ab}(\vn) \cD^{(+)}_{[a}\psi_{b]}(\vn) + \eta(\vn) \cD^{\dagger(-)}_{a} \psi_{a}(\vn) \nn \\
&&+ \frac{1}{2} \epsilon_{abcde}\chi_{de}(\vn+\hatbmu_a+\hatbmu_b+\hatbmu_c)\cD^{\dagger(-)}_{c}\chi_{ab}(\vn+\hatbmu_c) \Big)~,
\eea
when expanded up to second order in the fields using (\ref{eq:u}) and (\ref{eq:cu}), it becomes
\bea
S^{(2)}_F &\approx& \frac{1}{g^2}\sum_{\vk,a,b,c,d,e} \Tr \chi_{ab}(\vk)\Big[-f^*_a(\vk)\delta_{bc} + f^*_b(\vk)\delta_{ac}\Big]\psi_c(-\vk) + \eta(\vk)f_c(\vk)\psi_c(-\vk) \nn \\
&&+ \frac{1}{2} \epsilon_{abcde}\chi_{de}(\vk)e^{i(k_a+k_b)}f_c(\vk)\chi_{ab}(-\vk)\nn \\
\eea
Upon restricting the sum and rescaling the field $2\chi_{ab} \rightarrow \chi_{ab}$ the fermionic action becomes
\bea
S^{(2)}_F &\approx& \frac{1}{g^2} \sum_{\vk,a<b;c,d<e} \Tr \Big(\chi_{ab}(\vk)\Big[-f^*_a(\vk)\delta_{bc} + f^*_b(\vk)\delta_{ac}\Big]\psi_c(-\vk) + \eta(\vk)f_c(\vk)\psi_c(-\vk) \nn \\
&&+\frac{1}{2}
\epsilon_{abcde}\chi_{de}(\vk)e^{i(k_a+k_b)}f_c(\vk)\chi_{ab}(-\vk)\Big)
\eea
We can then write this action in the form of a matrix product
\bea
S^{(2)}_F &\approx& \frac{1}{g^2} \sum_{\vk} \left(
\Psi (\vk) \Psi(-\vk)
\right) \left(\frac{1}{4}\right)
\left(
\begin{tabular}{cc}
0 & $M(\vk)$ \\
$-M^T(\vk)$  & 0  \\
\end{tabular}
\right)
\left(
\begin{tabular}{c}
$\Psi(\vk)$ \\
$\Psi(-\vk)$ 
\end{tabular}
\right)
\nn \\ &=&
\frac{1}{4g^2}\sum_{\vk} \Phi(\vk) \mathcal{M} \Phi(\vk)
\label{matprod}
\eea
where $\Phi\equiv(\Psi(\vk),\Psi(-\vk))$ and $\Psi_i = (\eta, \psi_1,\dots,\psi_5,\chi_{12},\dots,\chi_{15},\dots,\chi_{45})$ and $M(\vk)$ is given in block matrix form
\begin{small}
\beq
\left(
\eta \ \psi_{a} \ \chi_{de}
\right)(\vk)
\left(
\begin{tabular}{ccc}
0 & $f_b(\vk)$ & 0 \\
$-f^*_a(\vk)$  & 0 & $ f_g(\vk)\delta_{ha}-f_h(\vk)\delta_{ga}$ \\
0 & $-f^*_d(\vk)\delta_{eb} + f^*_e(\vk)\delta_{db}$ & 
$\epsilon_{ghcde} q_{gh} f_c(\vk)$
\end{tabular}
\right)
\left(
\begin{tabular}{c}
$\eta$ \\
$\psi_{b}$ \\
$\chi_{gh}$
\end{tabular}
\right)(-\vk).\nn
\eeq
\end{small}
where $q_{gh} = e^{i(k_g + k_h)}$.
Notice that $M$ has the properties $M^T(\vk) = -M^*(\vk) = -M(-\vk)$. 

Using the property that $\sum_a \hatbmu_a = 0$ we can square the matrix to obtain
\beq
M^2(\vk) = -\sum_{a = 1}^5 |e^{ik_a} - 1|^2 {\mathbb I}_{16}  = -4\sum_{a = 1}^5 \sin^2\Big(\frac{k_a}{2}\Big) {\mathbb I}_{16}  = -\widehat{\vk}^2 {\mathbb I}_{16}~.
\eeq
Thus
\beq
M^{-1} = -\frac{1}{\widehat{\vk}^2}M
\eeq
and the inverse of the full fermion matrix is
\bea
\mathcal{M}^{-1}=-\frac{1}{\widehat{\vk}^2}  
\left(
\begin{tabular}{cc}
0 & $-M^T(\vk)$ \\
$M(\vk)$  & 0  \\
\end{tabular}
\right)~.
\eea
Then we can write the quadratic part of the fermionic action as
\bea
S_F^{(2)} &=& \frac{1}{4g^2}\sum_{\vk}\Tr\left[ \sum_{ij}\Phi_i(\vk)\mathcal{M}_{ij}(\vk)\Phi_j(\vk) \right] \nn \\ 
&=&\frac{1}{4g^2}\sum_{\vk} \sum_{ij,A,B} \Phi_i^A(\vk)\mathcal{M}_{ij}(\vk)\Phi_j^B(\vk) \Tr(T^A T^B) \nn \\
&=& \frac{1}{8g^2}\sum_{\vk} \sum_{ij,A,B} \Phi_i^A(\vk)\mathcal{M}_{ij}(\vk)\Phi_j^B(\vk) \delta_{AB}~,
\eea
where we have expanded the fermions as $\Phi = \Phi^AT^A$ and used $\Tr(T^AT^B) = \frac{1}{2}\delta_{AB}$. Thus we write the propagators as 
\beq
\langle \Phi_i^A(\vk)\Phi_j^B(\vk) \rangle = 2\mathcal{M}^{-1}_{ij}(\vk)\delta_{AB}
\eeq
or alternatively
\beq
\langle \Psi_i^A(\vk)\Psi_j^B(-\vk) \rangle = \frac{2}{\widehat{\vk}^2} M^{T}_{ij}(\vk)\delta_{AB}~.
\eeq
Notice that by switching the fields (with some relabeling) we have
\beq
\langle \Psi_i^A(-\vk)\Psi_j^B(\vk) \rangle = -\langle \Psi_j^B(\vk)\Psi_i^A(-\vk) \rangle = -\frac{2}{ \widehat{\vk}^2}
M^{T}_{ji}(\vk)\delta_{BA}= -\frac{2}{ \widehat{\vk}^2} M_{ij}(\vk)\delta_{AB}~.
\eeq
For a consistency check we replace $\vk$ with $-\vk$ and get
\beq
\langle \Psi_i^A(-\vk)\Psi_j^B(\vk) \rangle = \frac{2}{\widehat{\vk}^2} M_{ij}^{T}(-\vk)\delta_{AB}= -\frac{2}{ \widehat{\vk}^2} M_{ij}(\vk)\delta_{AB}~.
\eeq
We must also undo the earlier rescaling of the $\chi$ field giving a factor of $\frac{1}{2}$ in the $\psi \chi$ propagators and a factor of $\frac{1}{4}$ in the $\chi \chi$ propagators. It is also important to note that if we switch the direction of fermion flow in the propagators then we pick up an additional minus sign.
\FIGURE[t]{
  \includegraphics[scale=1.0]{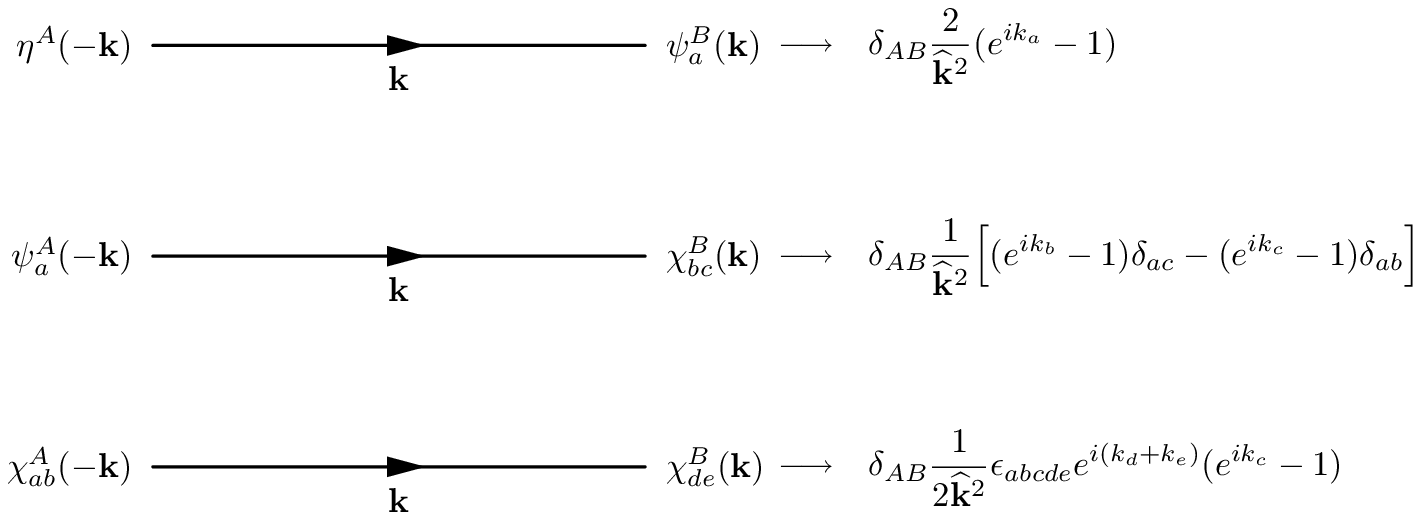}
  \caption{\label{fig:propagators}%
    The fermionic propagators.}
  }

%%%%%%%%%%%%%%%%%%%%%%%%%%%%%%%%%%%%%%%%%%%%%%%%%%%%%%%%%%%%%%%%%
\subsection{The Vertices}
%%%%%%%%%%%%%%%%%%%%%%%%%%%%%%%%%%%%%%%%%%%%%%%%%%%%%%%%%%%%%%%%%

Before we write down the expressions for vertices, let us further fix our conventions on the trace algebra. 
%For the generators $T^A$ of $SU(N)$ one has
%\beq
%T^AT^B = \frac{1}{2N}\delta_{AB}{\mathbb I}_N + \frac{1}{2}(d_{ABC} + if_{ABC}) T^C.
%\eeq
For the generators $T^A$ of $U(N)$ one has
\beq
T^AT^B =  \frac{1}{2}(d_{ABC} + if_{ABC}) T^C.
\eeq
where $d_{ABC}$ and $f_{ABC}$ are the symmetric and antisymmetric structure constants, respectively. This product formula is consistent with our previous trace convention $\Tr(T^AT^B) = \frac{1}{2}\delta_{AB}$ and in addition yields the results
\bea
\Tr(T^AT^BT^C) &=& \Tr\left( \frac{1}{2}(d_{ABD} + if_{ABD}) T^DT^C\right) \\ \nn 
&=& \frac{1}{2}(d_{ABD} + if_{ABD})\Tr[T^DT^C] \\ \nn
&=& \frac{1}{2}(d_{ABD} + if_{ABD})\frac{1}{2}\delta_{DC} \\ \nn
&=& \frac{1}{4}(d_{ABC} + if_{ABC}) = \frac{1}{4}\lambda_{ABC}~.
\eea
Since $f_{ABC}$ is antisymmetric and $d_{ABC}$ is symmetric it follows that
\beq
\lambda_{ACB} = \lambdabar_{ABC}~.
\eeq
%\bea
%\Tr(T^AT^BT^C) &=& \Tr\left(\frac{1}{2N}T^C + \frac{1}{2}(d_{ABD} + if_{ABD}) T^DT^C\right) \\ \nn 
%&=& \frac{1}{2}(d_{ABD} + if_{ABD})\Tr[T^DT^C] \\ \nn
%&=& \frac{1}{2}(d_{ABD} + if_{ABD})\frac{1}{2}\delta_{DC} \\ \nn
%&=& \frac{1}{4}(d_{ABC} + if_{ABC}) = \frac{1}{4}\lambda_{ABC}~.
%\eea
To extract expressions for the vertices we now return to the original gauge-fixed action for the theory  given by
\bea
S &=& \frac{1}{g^2}\sum_{\vn,a,b,c,d,e} \Tr \Big[\Big(\cD_a^{(+)}\cU_b(\vn)\Big)^{\dagger}\Big(\cD^{(+)}_a \cU_b(\vn)\Big) + \frac{1}{2} \Big(\cD^{\dagger(-)}_{a} \cU_{a}(\vn)\Big)^2 \nn \\
&&~~~~~~~~~~~~~~+ 2A_a(\vn)\partial^{(+)}_{a}\partial^{(-)}_bA_b(\vn)-
\Big(\chi_{ab}(\vn) \cD^{(+)}_{[a} \psi_{b]}(\vn) + \eta(\vn) \cD^{\dagger(-)}_{a} \psi_{a}(\vn)\nn \\
&&~~~~~~~~~~~~~~+\frac{1}{2}\epsilon_{abcde} \chi_{de}(\vn + \hatbmu_a + \hatbmu_b + \hatbmu_c) \cD^{\dagger(-)}_{c} \chi_{ab}(\vn + \hatbmu_c)\Big)\Big]~.
\eea
The last three terms of the action give rise to vertices between varying number of $\cA$'s and the fermions $\eta$, $\psi_a$, and $\chi_{ab}$. 
There are three vertices that arise at linear order in $\cA$:
\FIGURE[t]
  {
  \includegraphics[scale=1.0]{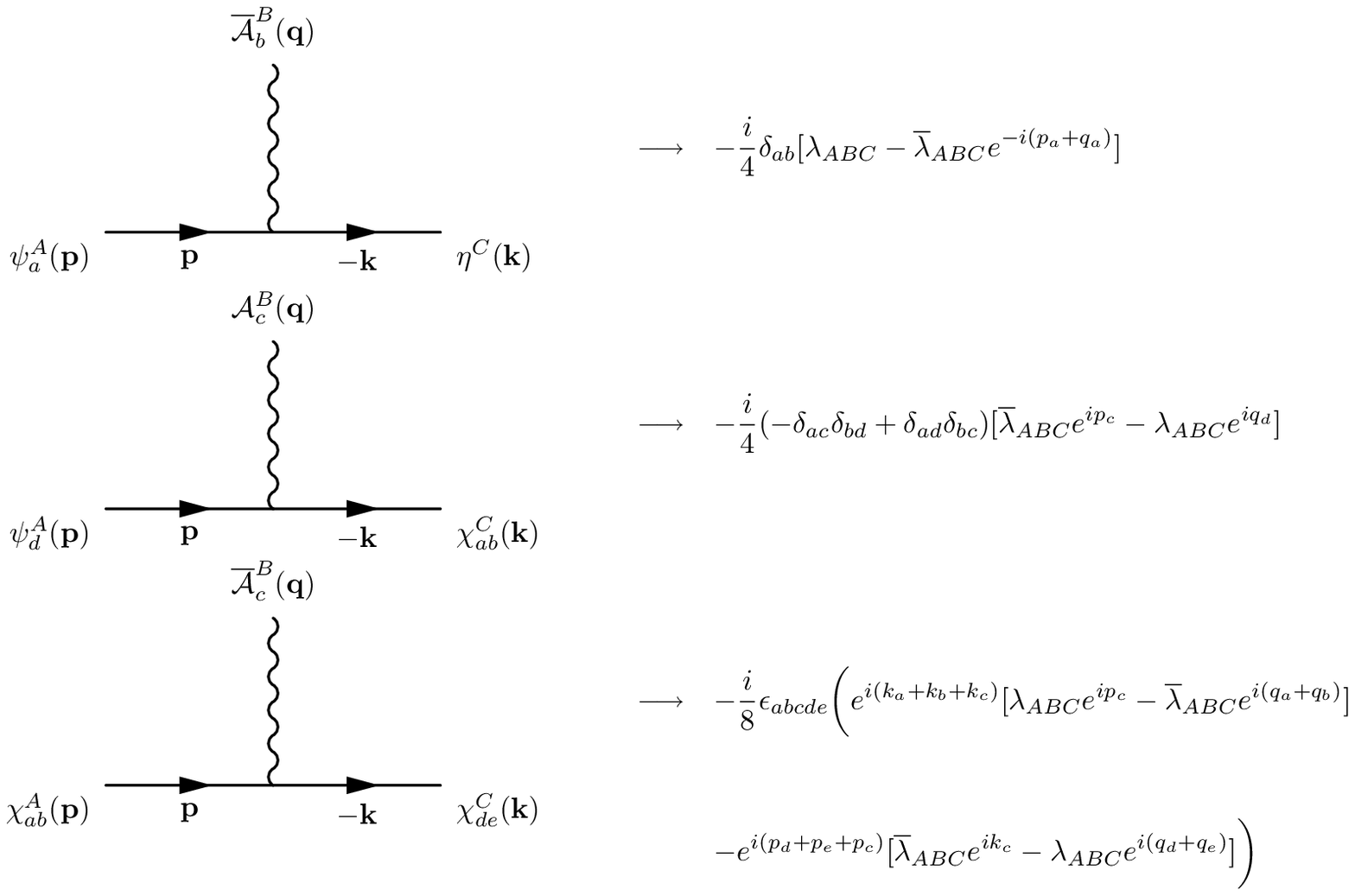}
  \caption{\label{fig:vertices}
    The vertices connecting fermions and complexified gauge fields.}
  }
\begin{itemize}
\item The $\psi\cAb\eta$ vertex
\bea
V_{\psi\cAb\eta} &=& -\sum_{\vn,a} \Tr\Big(\eta(\vn) \cD^{\dagger(-)}_{a} \psi_{a}(\vn)\Big) \nn \\
&=& -\sum_{\vn,a} \Tr ~\Big(\eta(\vn) \psi_a(\vn)\cU^{\dagger}_a(\vn) - \eta(\vn)\cU^{\dagger}_a(\vn - \hatbmu_a) \psi_a(\vn - \hatbmu_a)\Big) \nn \\
&=&-\sum_{\vn,\vk,\vq,\vp,a} \Tr ~e^{i(\vk+\vq+\vp)\cdot \vn}\Big(\eta(\vk) \psi_a(\vq)(-i)\cAb_a(\vp) - \eta(\vk)(-i)\cAb_a(\vp) e^{ip_a}\psi_a(\vq) e^{iq_a}\Big) \nn \\
&=& \sum_{\vk,\vq,\vp} \delta_{-\vk,\vq+\vp} \eta^C(\vk)\cAb_b^B(\vp)\psi_a^A(\vq)\left(\frac{i}{4}\right)\delta_{ab}[\lambda_{ABC} - \lambdabar_{ABC}e^{-i(p_a + q_a)}]~.
\eea
Thus the Feynman diagram contribution for this vertex is (add a minus since it comes from the first order term of $e^{-S}$)
\beq
V_{\eta\cAb\psi} = -\frac{i}{4}\delta_{ab}[\lambda_{ABC} - \lambdabar_{ABC}e^{-i(p_a + q_a)}]~.
\eeq
\item The $\psi\cA\chi$ vertex
\bea
V_{\psi\cA\chi} &=& -\sum_{\vn,a,b} \Tr \chi_{ab}(\vn) \cD^{(+)}_{[a}\psi_{b]}(\vn) \nn \\
&=& \sum_{\vn,a,b} \Tr \Big(-\chi_{ab}(\vn) \cD^{(+)}_a\psi_b(\vn)+\chi_{ab}(\vn) \cD^{(+)}_b\psi_a(\vn)\Big)\nn \\
&=& \sum_{\vn, a,b,c,d} (-\delta_{ac}\delta_{bd} + \delta_{ad}\delta_{bc})\Tr \left[\chi_{ab}(\vn)\Big(\cU_c(\vn)\psi_d(\vn + \hatbmu_c) - \psi_d(\vn)\cU_c(\vn + \hatbmu_d)\Big)\right]\nn \\
&=& \sum_{\vk,\vq,\vp, a,b,c,d}\delta_{-\vk,\vq+\vp} (-\delta_{ac}\delta_{bd} + \delta_{ad}\delta_{bc})\chi^C_{ab}(\vk)\cA^B_c(\vq)\psi^A_d(\vp)\frac{i}{4}[\lambdabar_{ABC}e^{ip_c} - \lambda_{ABC}e^{iq_d}]~.~~~~~~~~~
\eea
The vertex is given by 
\beq
V_{\chi\cA\psi} = -\frac{i}{4}(-\delta_{ac}\delta_{bd} + \delta_{ad}\delta_{bc})[\lambdabar_{ABC}e^{ip_c} - \lambda_{ABC}e^{iq_d}]~.
\eeq
\item The $\chi\cAb\chi$ vertex
\bea
V_{\chi\cAb\chi} &=& -\frac{1}{2}\sum_{\vn,a,b,c,d,e} \Tr \epsilon_{abcde}\chi_{de}(\vn+\hatbmu_a+\hatbmu_b+\hatbmu_c)\cD^{\dagger(-)}_{c}\chi_{ab}(\vn+\hatbmu_c)\nn \\
&=&-\frac{1}{2}\sum_{\vn,a,b,c,d,e} \Tr \epsilon_{abcde}\chi_{de}(\vn+\hatbmu_a+\hatbmu_b+\hatbmu_c)\Big( \chi_{ab}(\vn+\hatbmu_c) \cU^{\dagger}_c(\vn)- \cU_c^{\dagger}(\vn + \hatbmu_a + \hatbmu_b) \chi_{ab}(\vn)\Big)\nn \\
&=& \frac{1}{2}\sum_{\vk,\vp,\vq,a,b,c,d,e} \delta_{-\vk,\vq+\vp}\epsilon_{abcde}\chi^C_{de}(\vk)\cAb^B_c(\vq)\chi^A_{ab}(\vp) \bigg(e^{i(k_a+k_b+k_c)}\frac{i}{4}[\lambda_{ABC}e^{ip_c} - \lambdabar_{ABC}e^{i(q_a + q_b)}]\nn \\
&& -e^{i(p_d+p_e+p_c)}\frac{i}{4}[\lambdabar_{ABC}e^{ik_c} - \lambda_{ABC}e^{i(q_d + q_e)}] \bigg)~.\nn
\eea
The vertex is given by (taking into account both possible contractions with external propagators)
\beq
V_{\chi\cAb\chi} = -\frac{i}{8}\epsilon_{abcde}\bigg(e^{i(k_a+k_b+k_c)} [\lambda_{ABC}e^{ip_c} - \lambdabar_{ABC}e^{i(q_a + q_b)}] - e^{i(p_d+p_e+p_c)}[\lambdabar_{ABC}e^{ik_c} - \lambda_{ABC}e^{i(q_d + q_e)}] \bigg)~.
\eeq
\end{itemize}
%%%%%%%%%%%%%%%%%%%%%%%%%%%%%%%%%%%%%%%%%%%%%%%%%%%%%%%%%%%%%%%%%
\section{One Loop Diagrams for the Renormalized Fermion Propagators}
%%%%%%%%%%%%%%%%%%%%%%%%%%%%%%%%%%%%%%%%%%%%%%%%%%%%%%%%%%%%%%%%%
\label{sec:1-loop}
Using these propagators and vertices it is straightforward to see that the renormalized fermion propagators receive contributions from the following four {\em amputated} diagrams. 
\begin{itemize}
\item The amputated $\eta\psi$ diagram. 
We have an $\cAb\cA$ propagator, a $\psi\chi$ propagator, an $\eta\cAb\psi$ vertex, and a $\chi\cA\psi$ vertex. Using the expressions above we have
\bea
I_{\eta \psi}(\vp) &=& \sum_{\vk, \vq} \sum_{BC} \sum_{abc} \delta_{-\vp,\vk+\vq}\Big[\frac{1}{\widehat{\vk}^2} [(e^{ik_b}-1)\delta_{ac}-(e^{ik_c} - 1) \delta_{ab}]\Big]\cdot\Big[ \frac{1}{\widehat{\vq}^2}\Big]\cdot \Big[\frac{i}{4}[\lambda_{ABC} - \lambdabar_{ABC}e^{i(k_a+q_a)}]\Big]\nn \\
&&\cdot \Big[ \frac{i}{4}(-\delta_{ba}\delta_{cd} +  \delta_{bd}\delta_{ca})[\lambdabar_{BCD}e^{-ip_a} - \lambda_{BCD} e^{iq_d}]\Big]~.
\eea
\item The first amputated $\psi\chi$ diagram.
We have an $\cA \cAb$ propagator, a $\chi\chi$ propagator, a $\psi\cA\chi$ vertex, and a $\chi\cAb\chi$ vertex. 
\bea
I^{1}_{\psi\chi}(\vp) &=& \sum_{\vk,\vq} \sum_{bcdefm} \sum_{BC}\Big[ \frac{1}{2\widehat{\vk}^2}\epsilon_{bcmef}e^{i(k_e + k_f)}(e^{ik_m} - 1)\Big] \nn \\
&&\cdot \Big[ \frac{1}{\widehat{\vq}^2}\Big]\cdot \Big[ -\frac{i}{4}(-\delta_{bd}\delta_{ca} + \delta_{ba} \delta_{cd})[\lambdabar_{ACB}e^{ip_d}-\lambda_{ACB}e^{-iq_a}]\Big]\nn \\
&&\cdot\bigg[\frac{i}{8}\epsilon_{efdgh} \Big(e^{ik_{(d+g+h)}}[\lambdabar_{BCD}e^{-ip_d} - \lambda_{BCD}e^{i(q_g + q_h)}]\Big]\nn \\
&& -e^{-ip_{(d+e+f)}}[\lambda_{BCD}e^{ik_d} - \lambdabar_{BCD}e^{i(q_e + q_f)}]\Big)\bigg]~.
\eea
\item The second amputated $\psi\chi$ diagram has an $\cAb\cA$ propagator, an $\eta\psi$ propagator, a $\psi\cAb\eta$ vertex, and a $\psi\cA\chi$ vertex. This yields
\bea
I^{2}_{\psi\chi}(\vp) &=& \sum_{\vk,\vq} \sum_{bc} \sum_{BC} \Big[\frac{2}{\widehat{\vk}^2}(e^{ik_c} - 1)\Big]\cdot \Big[ \frac{1}{\widehat{\vq}^2}\Big] \cdot \delta_{ab} \Big[-\frac{i}{4}[\lambda_{ACB} - \lambdabar_{ACB}e^{-i(p_a-q_a)}]\Big]\nn \\
&&\cdot \Big[-\frac{i}{4}(-\delta_{db}\delta_{ec} + \delta_{dc}\delta_{eb})[\lambda_{DCB}e^{ik_b} -\lambdabar_{DCB}e^{iq_c}]\Big]~.
\eea
\item The amputated $\chi\chi$ diagram. 
It has a $\cAb\cA$ propagator, a $\chi\psi$ propagator, a $\chi\cAb\chi$ vertex, and a $\psi\cA\chi$. 
\bea
I_{\chi\chi}(\vp) &=&\sum_{\vk,\vq} \sum_{cdef} \sum_{BC} \delta_{\vk+\vq-\vp, 0} \Big[\frac{1}{\widehat{\vk}^2}[(e^{-ik_e} - 1)\delta_{fd} - (e^{-ik_d} - 1)\delta_{fe}]\Big]\cdot\Big[\frac{1}{\widehat{\vq}^2}\Big]\nn \\
&&\cdot\bigg[-\frac{i}{8}\epsilon_{abcde} \Big(e^{-ik_{(a+b+c})}[\lambda_{ACB}e^{ip_c} - \lambdabar_{ACB}e^{-i(q_a + q_b)}]\nn \\
&&-e^{ip_{(c+d+e})}[\lambdabar_{ACB}e^{-ik_c}    - \lambda_{ACB}e^{-i(q_d + q_e)}]\Big)\bigg]\nn \\
&&\cdot\Big[-\frac{i}{4}(-\delta_{gc}\delta_{hf} + \delta_{gf}\delta_{hc})[\lambdabar_{BCD}e^{ik_c} - \lambda_{BCD}e^{iq_f}]\Big]~.
\eea
\end{itemize}
\FIGURE[t]
  {
  \includegraphics[scale=1.0]{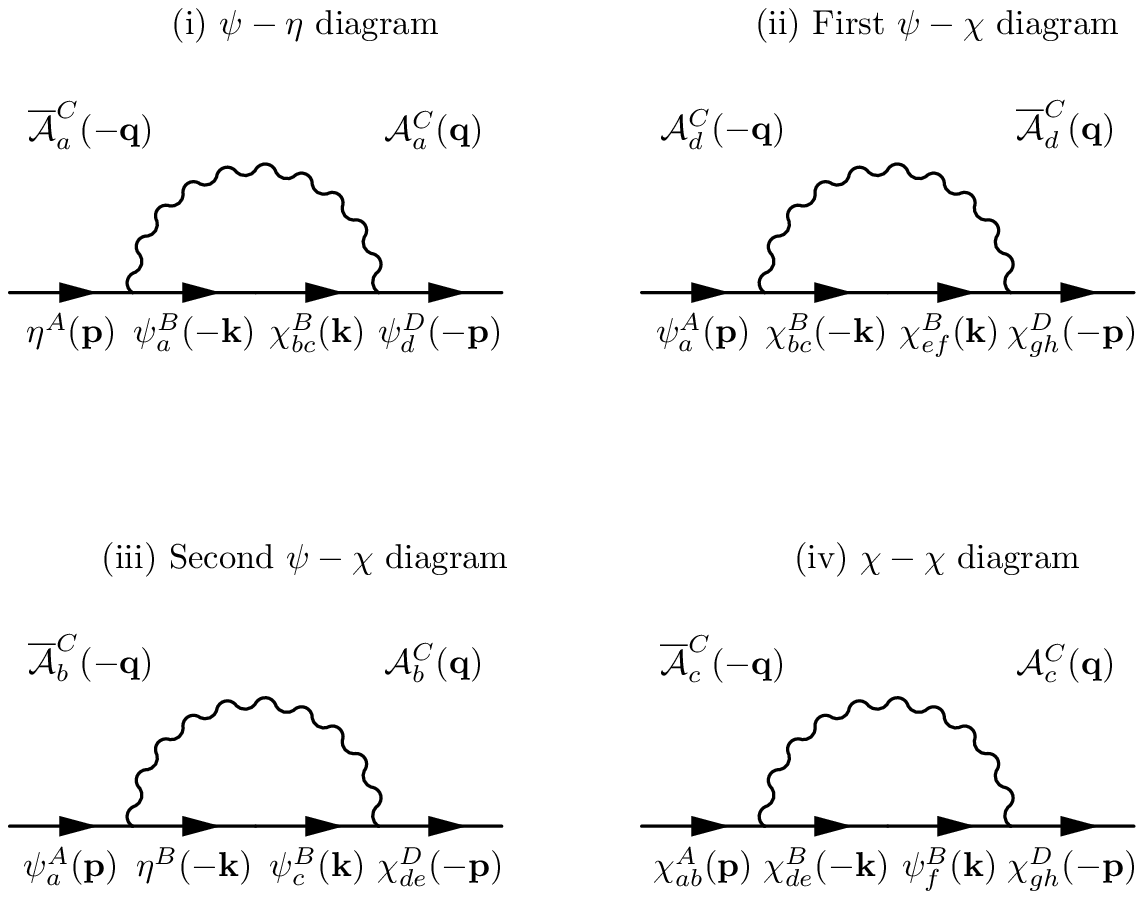}
  \caption{\label{fig:one-loop}
    The one loop diagrams of fermions and complexified gauge fields.}
  }
In appendix A we show that the contributions of these diagrams all vanish in the limit $p \rightarrow 0$ indicating that mass counterterms are absent in the lattice theory at one loop. In our general argument of section~\ref{sec:general} we argued that the only dangerous mass term involved a coupling of $\eta$ and $\psi_a$. We now see that this term does not arise at one loop. In the next section we will show that this feature persists to all orders and thus our general conclusion will be that {\it no mass counterterms are needed at any finite order of perturbation theory}.

%%%%%%%%%%%%%%%%%%%%%%%%%%%%%%%%%%%%%%%%%%%%%%%%%%%%%%%%%%%%%%%%%
\section{Effective Action}
%%%%%%%%%%%%%%%%%%%%%%%%%%%%%%%%%%%%%%%%%%%%%%%%%%%%%%%%%%%%%%%%%

In this section we will compute the partition function of the lattice theory in one loop order around an arbitrary classical vacuum state in which the fermions vanish and the bosonic fields correspond to constant commuting matrices. To start we expand the fields around such a constant commuting background,
\beq
\cU_a(\vn) = \cU_a + i \cA_a(\vn), \quad \cU^{\dagger}_a(\vn) = \cU^{\dagger}_a - i \cAb_a(\vn)
\eeq
Choosing the gauge $\alpha=1/2$, the quadratic part of the bosonic action then takes the form
\beq
S_B = -2 \sum_{\vn, a, b} \Tr \cA_b(\vn) \cD_a^{\dagger(-)} \cD_a^{(+)} \cA_b(\vn)~.
\eeq
Here the covariant derivatives depend on the constant 
commuting classical background $\[\cU_a, \cU^{\dagger}_a\]=0$. 
After integration over the fluctuations in the bosonic
fields one finds the bosonic contribution to the one loop
partition function is given by
\beq {\rm det}^{-5}( \cD_a^{\dagger(-)} \cD_a^{(+)} )\eeq

The gauge fixing functional \myref{gfixf} leads to the quadratic ghost action
\beq
S_G = \sum_{\vn,a} \Tr \cbar\, \cD_a^{\dagger (-)} \cD_a^{(+)} c~.
\eeq
The quadratic fermionic part of the action is given by the corresponding terms in \myref{eq:final-action}, except that now the covariant derivatives depend only on the background fields.

Since the background is constant, we can pass to momentum space in which the action separates into terms for each mode $\vk$.  The $16 \times 16$ fermion matrix $M(\vk)$ for the mode $\vk$ then can be shown (using MAPLE to compute the determinant) to satisfy
\beq
\det M(\vk) = \det ( \cD_a^{\dagger(-)}(\vk) \cD_a^{(+)}(\vk) )^8~.
\eeq
Going back to position space, and taking into account the fact that there is a double counting of modes in the matrix form \myref{matprod}, we obtain \beq
{\rm Pf} ({\cal M}) = {\rm det}^4( \cD_a^{\dagger(-)} \cD_a^{(+)} )~.
\eeq
The ghosts add another factor of $\det (\cDb_a^{\dagger(-)} \cD_a^{(+)})$, which is just what is needed to cancel the bosonic contribution given
earlier.  

In conclusion, we have shown that the one loop
effective action of the lattice theory
obtained by expanding about an arbitrary
point in the classical moduli space is identically zero.
Thus, as for the continuum,
the moduli space is not lifted in this analysis and hence 
there can be no boson or fermion masses at one loop. Furthermore, we expect that we can extend this analysis to all loops since the partition function of the lattice theory is a topological invariant and hence can be computed exactly in the semi-classical approximation (see Appendix C). Indeed, Matsuura uses similar arguments to show that the vacuum energy of supersymmetric lattice theories with four and eight supercharges remains zero to all orders in the coupling \cite{Matsuura:2007ec}. The calculation presented here extends this to the case of sixteen supercharges\footnote{Notice that in this calculation we have not included any mass terms that would guarantee the stability of the initial classical vacuum state we have chosen to expand around. We have also ignored a potential sign problem associated with the replacement of a Pfaffian with a square root of a determinant. Nevertheless we expect the result to be robust; the existence of an exact supersymmetry should ensure that the object we are computing is a lattice regularized Witten index and hence independent of both coupling constant and background field.}. 
Thus we conclude that boson and scalar masses remain zero to all orders in the coupling constant. This implies that the fermions
also remain massless which is consistent with our explicit one loop calculation.

At this point we have derived expressions for the amputated one loop diagrams that contribute to the renormalization of the three twisted fermion propagators. This is sufficient to calculate $\alpha_1,\alpha_2$ and $\alpha_4$ that appear in the general action 
\bea
{\cal L} &=& \frac{1}{g^2}\sum_{\vn,a,b,c,d,e} ~\Big\{ \cQ ~\Tr \Big[-i\alpha_1 \chi_{ab} \cD^{(+)}_a \cU_b(\vn) -i \alpha_2 \eta(\vn) \cD^{\dagger(-)}_a \cU_a(\vn) +\frac{\alpha_3}{2}\eta(\vn)d(\vn)\Big]\nn \\
&& - \frac{\alpha_4}{2} \Tr \epsilon_{abcde} \chi_{de}(\vn + \hatbmu_a + \hatbmu_b + \hatbmu_c) \cD^{\dagger(-)}_{c} \chi_{ab}(\vn + \hatbmu_c)\Big\}~.
\eea
However the coefficient $\alpha_3$ requires further work. One simple
way to extract it is via a computation of the renormalized auxiliary
boson propagator which we turn to in the next section.

\section{One Loop Diagrams for the Auxiliary Field Propagator}

We have shown that the off-shell form of the bosonic action is given by
\beq
S_B = \sum_{\vn,a,b} \Tr \Big( \cF_{ab}^{\dagger}(\vn) \cF_{ab}(\vn)  -\frac{i}{g}d(\vn) \cD^{\dagger(-)}_a \cU_a(\vn) + \frac{1}{2} d^2(\vn)\Big)~,
\eeq 
where $\cF_{ab}(\vn) = -\frac{i}{g}\cD_a^{(+)}\cU_b(\vn)$. 

\begin{figure}
\bec
\includegraphics[scale=1.1]{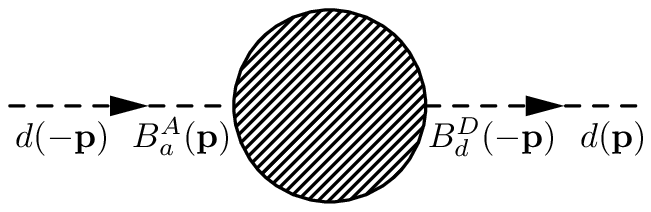}
\caption{\label{Bbubble}The generic diagram contributing to the renormalized $d$ propagator}
\eec
\end{figure}

In our previous computation of the fermion diagrams we integrated out the field $d$ to give an on shell action defined just in terms of the complex gauge link fields $\cU_a$ and $\cU^{\dagger}_a$. In this section we will not do this but instead focus on a computation of the renormalized propagator for the $d$ field. The Feynman rules for the fermions will be identical to our previous scheme but the boson propagators will change and so we need to recompute those propagators in this off shell scheme. We proceed in the standard fashion by expanding the link field $\cU_a(\vn)$ 
\beq
\cU_a (\vn) =  {\bf 1} + ig\cA_a(\vn), ~~\cU^{\dagger}_a (\vn) = {\bf 1} - ig\cAb_a(\vn).
\eeq
and using the same lattice gauge-fixing term as before 
\beq
S_{GF}[A] = -\frac{1}{\alpha} \sum_{\vn,a} \Tr (\partial^{(-)}_{a}A_a(\vn))^2,
\eeq
we find the momentum space form
\beq
S_{GF}[A] =\frac{1}{\alpha} \sum_{\vk,a,b} \Tr A_a(\vk) f^*_a(\vk)f_b(\vk) A_b(-\vk)~.
\eeq
It is convenient in this calculation to work with the real and imaginary parts of the complex gauge field explicitly. Thus 
\beq 
\cA_a=A_a+iB_a
\eeq
The gauge-fixed bosonic action on the lattice to quadratic order in fields, with the choice $\alpha = \frac{1}{2}$, is then
\bea
S^{(2)}_B &=&\sum_{\vk,a,b} \Tr 2A_a(\vk)\Big[\delta_{ab}f_c(\vk)f^*_c(\vk)\Big]A_b(-\vk)+2B_a(\vk)\Big[\delta_{ab}f_c(\vk)f^*_c(\vk)-f^*_a(\vk)f_b(\vk)\Big]B_b(-\vk) \nn \\
&&-2id(\vk) f_a(\vk) B_a(-\vk) +\frac{1}{2} d(\vk) d(-\vk)
\label{offshellA}
\eea 
We see that the $d-B_a$ system decouples from $A_a$ to this order. 
Its action is given by
\bea
S^{(2)}_B [d, B_a]&\sim& \sum_{\vk,a,b} \Tr 2B_a(\vk)\Big[\delta_{ab}f^*_c(\vk)f_c(\vk)-f^*_a(\vk)f_b(\vk)\Big]B_b(-\vk)\nn \\
&&-2id(\vk) f_a(\vk) B_a(-\vk) +\frac{1}{2} d(\vk) d(-\vk)
\eea 
or in matrix form
\bea
\left(\begin{tabular}{cc}$d$ & $B_a$ \end{tabular}\right)(\vk)\left(
\begin{tabular}{cc}
$\frac{1}{2}$ & $-if_b(\vk)$ \\
$-if^*_a(\vk)$  & $M_{ab}(\vk)$
\end{tabular}
\right)
\left(
\begin{tabular}{c}
$d$ \\
$B_b$ 
\end{tabular}
\right)(-\vk)
\eea
where $M_{ab}(\vk) = 2[\delta_{ab}\sum_c f_c(\vk)f^*_c(\vk) - f^*_a(\vk)f_b(\vk)]$.
Using standard identities for the inverse of a partitioned
matrix we find
\beq
M^{-1} = \left(
\begin{tabular}{cc}
$\frac{1}{2}$ & $-if_b(\vk)$ \\
$-if^*_a(\vk)$  & $M_{ab}(\vk)$
\end{tabular}
\right)^{-1} = \frac{1}{\sum_c f_c(\vk)f^*_c(\vk)}\left(
\begin{tabular}{cc}
0 & $if_b(\vk)$ \\
$if^*_a(\vk)$  & $\frac{1}{2}\mathbf{1}_5$
\end{tabular}
\right)
\eeq
We have $\sum_c f_c(\vk)f^*_c(\vk) = 4 \sum_c \sin^2\Big(\frac{\vk_c}{2}\Big)$ and as before we define $\widehat{\vk}^2 \equiv 4 \sum_c \sin^2\Big(\frac{\vk_c}{2}\Big)$. Thus the lattice propagators are
\bea
\langle d^A(\vk)d^B(-\vk) \rangle &=& 0 \\
\langle d^A(\vk)B^B_a(-\vk) \rangle &=& i\delta_{AB}\frac{(e^{-ik_a}-1)}{\widehat{\vk}^2} \\
\langle B^A_a(\vk)B^B_b(-\vk) \rangle &=& \delta_{ab}\delta_{AB}\frac{1}{2\widehat{\vk}^2}
\eea
From eqn~\ref{offshellA} the propagator for the $A$ field is also
\beq
\langle A^A_a(\vk)A^B_b(-\vk) \rangle = \delta_{ab}\delta_{AB}\frac{1}{2\widehat{\vk}^2}~.
\eeq
Notice that the field $d$ is non-propagating at tree level. 
Using these propagators and those derived earlier for the fermions and
ghosts
we can now write down the generic Feynman diagram
contributing to a renormalization of the auxiliary boson propagator. It is
shown in figure~\ref{Bbubble} and represents the set of amputated 
diagrams possessing two external B field legs. These combine with the external
$<dB>$ propagators derived above
to yield the renormalized propagator for the auxiliary
field $d$. Notice that the vanishing of the tree level $<dd>$ propagators
ensures that no amputated diagrams with 2 $d$ field external legs
contribute.
The set of all such lattice Feynman diagrams is shown below and
corresponds to a subset of the $B$ field vacuum polarization
diagrams.
\begin{figure}
\bec
\includegraphics[scale=1.0]{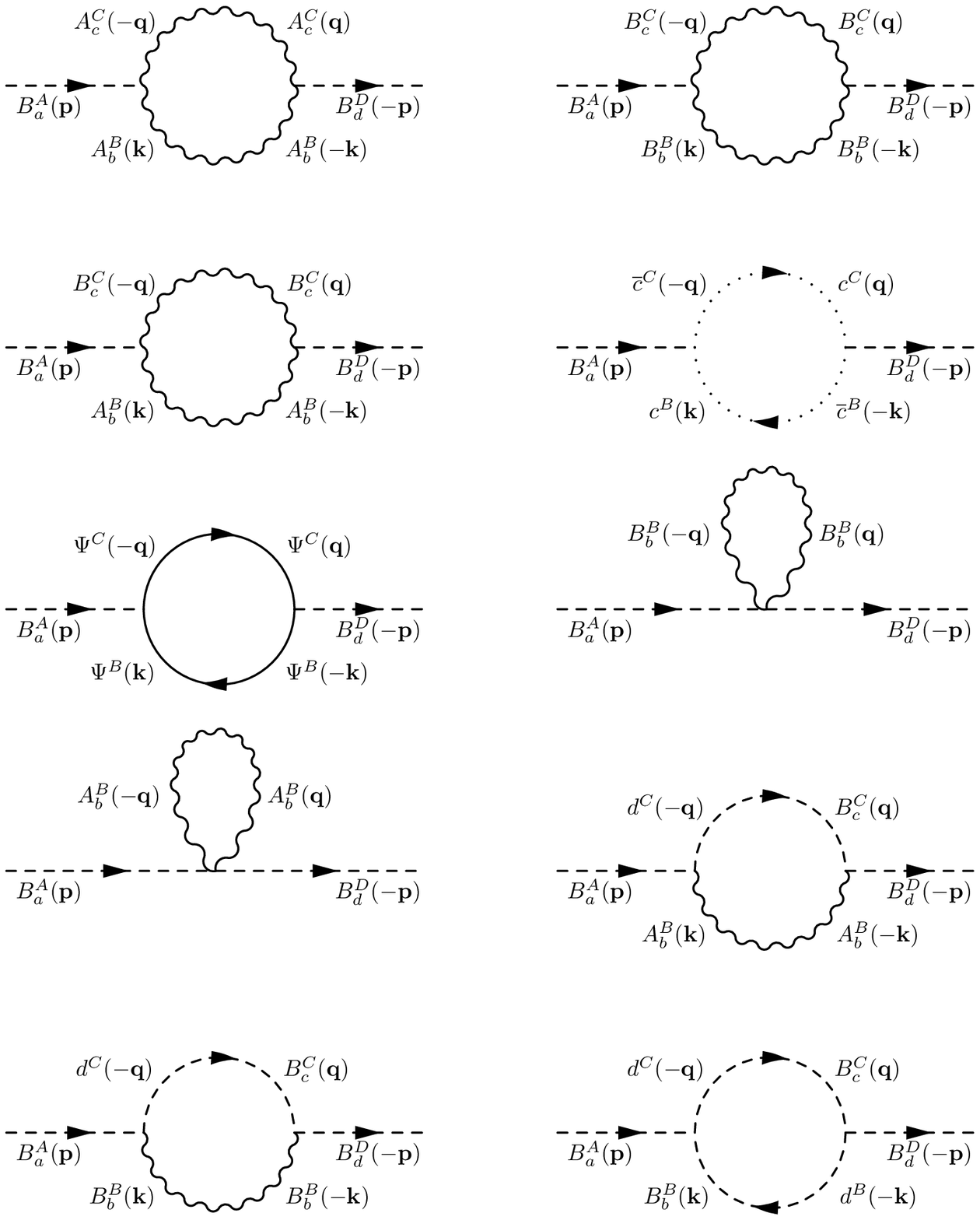}
\caption{\label{AllBbubbles} Set of all lattice amputetated 
Feynman diagrams contributing the renormalized $d$ propagator}
\eec
\end{figure}
It is important
to notice that almost all these diagrams appear in the continuum off shell
twisted theory; the exceptions are just the diagrams containing a
$BBd$ vertex which corresponds to the lattice vertex
\beq
V_{dBB} = \langle d^A(-\vk-\vq)B^B_a(\vk)B^C_b(\vq) \rangle = \frac{i}{2}\delta_{ab}(\lambda_{ABC}+\lambdabar_{ABC})(1-e^{-i(k_a+q_a)})~.
\eeq
Clearly this vertex vanishes as the lattice spacing is sent to zero and
hence this diagram does not contribute to the divergent piece
in the $<dd>$ propagator
at this order of perturbation theory.

Hence we are left with a set of diagrams which correspond to those
of the equivalent continuum theory at one loop order. This fact can
be exploited later to allow us to argue that the leading logarithmic
divergences of the lattice theory are shared with the continuum theory.
Anticipating this we will not
write down explicit expressions for these
amputated lattice diagrams in this section.

%%%%%%%%%%%%%%%%%%%%%%%%%%%%%%%%%%%%%%%%%%%%%%%%%%%%%%%%%%%%%%%%%
\section{Divergence Structure of the One Loop Diagrams}
%%%%%%%%%%%%%%%%%%%%%%%%%%%%%%%%%%%%%%%%%%%%%%%%%%%%%%%%%%%%%%%%%

At this point we have derived expressions for the amputated one loop diagrams that determine the renormalization of three fermion propagators and also the
set of Feyman graphs needed to
renormalize the auxiliary bosonic field propagator. In principle, this input
will allow us to determine all four coefficients $\alpha_i$ appearing in the renormalized action eqn.~\ref{eq:general-form}. Of course the question of how much fine tuning is required to regain full supersymmetry is determined by the parts of these expressions which diverge as the lattice spacing is sent to zero. We must therefore evaluate the expressions for the one loop integrals as the lattice spacing tends to zero.

First, let us discuss the diagrams contributing to the fermion
propagators.
We have shown in Appendix B that the three fermion amplitudes all vanish for vanishing external momentum which is consistent with our effective action computation showing that no fermionic mass terms can be generated perturbatively. Reisz's power counting theorem \cite{Reisz:1988a}-\cite{Reisz:1988d} shows us that we cannot simply take the na\"ive continuum limit of the expressions for the amputated one loop diagrams as they have a na\"ive degree of divergence of 1.  However we can use a trick due to \cite{Kawai:1981} and detailed in \cite{Capitani:2002} to extract the leading divergences.

We split the integral $I(\mathbf{p})$ into two pieces as follows:
\begin{eqnarray}
\lim_{a \rightarrow0} I(\mathbf{p})&=& \lim_{a \rightarrow0}\left[ I(\mathbf{p}) -I(\mathbf{0})-\sum_b p_b \frac{\partial I}{\partial p_b} \bigg|_{\mathbf{p}=\mathbf{0}} \right] +\lim_{a \rightarrow0}\left[I(\mathbf{0})+\sum_b p_b \frac{\partial I}{\partial p_b}\bigg|_{\mathbf{p}=\mathbf{0}}\right]
\end{eqnarray}
The first term in square brackets can now be evaluated in the na\"ive continuum limit and contains no divergence. The second term contains the divergence but contains no external momenta in the integrand which simplifies its evaluation on the lattice. In addition we know that $I(\mathbf{0})$ vanishes for each of our diagrams so the calculation becomes simpler still.

We will find that
the resulting expressions have logarithmic divergences of the form
$\ln{\mu a}$ where $\mu$ is a small mass parameter used to regulate
the behavior of the integrand close to the origin of momentum
space and $a$ the lattice
spacing\footnote{We will only consider the case of infinite lattice
  size which reduces all lattice sums in momentum space to
  integrals.}. 

One obvious way to proceed is simply to numerically evaluate the
integral for a variety of regular masses $\mu$ and extract the logarithmic
divergence and any constant contributions using a fitting procedure.
However, if we are only interested in the leading log divergences there
is a simpler approach detailed in the next section in which a na\"ive
continuum limit can be taken and the expressions evaluated using, for
example, dimensional regularization.

%We then use dimensional regularisation to calculate the na\"ive
%continuum limit integrals giving
%\bea {\label{eqn:logterms}}
%\lim_{a \rightarrow 0} I(\mathbf{p})&=& \mathbf{L}\left[ b_C \log (p^2 +
%  \mu^2)a^2 - b_C \log \mu^2 a^2 \right]
%+\mathbf{L}^{\prime} b_L \log \mu^2 a^2
%\eea
%$\mathbf{L}$ and $\mathbf{L}^\prime$ are the Lorentz factors
%associated with the two integrals, $b_C$ denotes the coefficient of the
%$\log$ term obtained through dimensional regularisation and $b_L$ the
%term obtained from the lattice integral. Note that the term in the
%square brackets does not have a divergence in it (by construction) and
%so the two $\log$ terms must have the same coefficient. 

In the next section we give an example of this procedure for the amputated $\eta \psi$ diagram and show how to extract similar results for the remaining fermion self energy diagrams. We will also see that the same procedure allows us
argue that the leading log divergent contribution to $\alpha_3$ is also
equal to its value in the continuum theory. 

\subsection{The Amputated Fermion Diagrams}
We start with our simplified expression for $I_{\eta \psi_d}(\mathbf{p})$ given in Appendix A
\bea
I_{\eta \psi_d}(\mathbf{p})&=& \int \frac{d^4 \mathbf{q}}{(2 \pi)^4}  \sum_{BC} \left[ \frac{1}{8 \widehat{(\mathbf{p}-\mathbf{q})}^2 
\widehat{\mathbf{q}}^2} (1-e^{i(p-q)_d}) \right]\bigg[-\sum_{a \neq d} \big[d_{ABC} d_{BCD} (e^{-ip_a} -e^{iq_d} -1 + e^{ip_a +iq_d})\nn \\ 
&&+f_{ABC} f_{BCD} (e^{-ip_a} +e^{iq_d} +1 + e^{ip_a +iq_d})\big]\bigg]~.
\eea
As a first step we need to calculate the derivative of the diagram (re-inserting the lattice spacing $a$ and the infra-red cutoff $\mu$)
\bea
\frac{\partial I_{\eta \psi_d}(\mathbf{p})}{\partial p_b}
\bigg|_{\mathbf{p}=\mathbf{0}}
 &=& 
\int_{\frac{-\pi}{a}}^{\frac{\pi}{a}} \frac{d^4 \mathbf{q}}{(2 \pi)^4}
\frac{-2a^4 \sin aq_b} { (\widehat{\mathbf{q}}^2+\mu^2 a^2)^3 }
( 1 - e^{-iaq_d}) 
f_{ABC} f_{BCD} (
1 +  e^{iaq_d})
\nn \\ &&
+
\int_{\frac{-\pi}{a}}^{\frac{\pi}{a}} \frac{d^4 \mathbf{q}}{(2 \pi)^4}
\frac{-a^3}{(\widehat{\mathbf{q}}^2 + \mu^2 a^2)^2} 
(  - ia \delta_{db} e^{-iaq_d}) 
 f_{ABC} f_{BCD} (
 1 +  e^{iaq_d})
\nonumber \\ &&
+
\int_{\frac{-\pi}{a}}^{\frac{\pi}{a}} \frac{d^4 \mathbf{q}}{(2 \pi)^4}
\frac{-a^3}{8(\widehat{\mathbf{q}}^2 +\mu^2 a^2)^2} 
( 1 - e^{-iaq_d}) 
\nonumber \\
&& \times \sum_{a \neq d}
(d_{ABC} d_{BCD} +f_{ABC} f_{BCD}) \delta_{ab}(-ia)(
 1 - e^{iaq_d})~.
\eea
A further simplification now occurs; if we are only interested in the
leading $\log \mu a$ coefficient we can evaluate this integral in a
small $q$ region around zero. This is because the contribution of the
integrand to the $\log \mu a$ coefficient comes only from small
$q$. Furthermore in the region $q\to 0$ the propagators and vertices
inside the integral will approach their continuum counterparts and hence the logarithmic divergence can be extracted by replacing the lattice integrals by their na\"ive continuum limit. Note that this only works for the coefficient of the $\log$ - we must evaluate the integral numerically (and then fit) in order to extract the constant terms. This (longer)
calculation is in progress and will be published in a future paper. Thus we find:
\begin{eqnarray}
\lim_{a \rightarrow 0}
\frac{\partial I_{\eta \psi_d}(\mathbf{p})}{\partial p_b}
 \bigg|_{\mathbf{p}=\mathbf{0}}
 &\sim& 
\int_{-\infty}^{\infty} \frac{d^4 \mathbf{q}}{(2 \pi)^4}
\frac{- 4iq_b q_d} { (q^2+\mu^2 )^3 } 
f_{ABC} f_{BCD}
\nn \\ &&
+
\int_{-\infty}^{\infty} \frac{d^4 \mathbf{q}}{(2 \pi)^4}
\frac{2i}{(q^2 + \mu^2 )^2} \delta_{db}  
 f_{ABC} f_{BCD}~.
\end{eqnarray}
Note that we cannot just set the first term in this expression to zero as $\widehat{e}_d$ and $\widehat{e}_b$ are not orthogonal to each other, instead we have
\begin{eqnarray}
\int d^d\mathbf{q} \frac{q_b q_d}{(q^2 +\mu^2)^3}
&=& e_b^\mu e_d^\nu
\int d^d\mathbf{q} \frac{q^\mu q^\nu }{(q^2 +\mu^2)^3}
\nn \\ 
&=& \widehat{e}_b \cdot \widehat{e}_d
\int d^d\mathbf{q} \frac{q^2}{d(q^2 +\mu^2)^3}~.
\end{eqnarray}
Then $ \widehat{e}_b \cdot \widehat{e}_d=\delta_{bd} - \frac{1}{5}$. We use dimensional regularization and the fact that $\sum_b p_b =0$ to evaluate the resulting integrals getting
\begin{eqnarray}
I_{\eta \psi_d}(\mathbf{p}) &\sim& 
\sum_b p_b \frac{\partial I_{\eta \psi_d}(\mathbf{p})}{\partial p_b}
 \bigg|_{\mathbf{p}=\mathbf{0}}
\nn \\ &\sim& -\frac{i}{8 \pi^2}
p_d f_{ABC}f_{BCD} \log \mu a~.
\end{eqnarray}
Note that we have inserted the cutoff $\frac{1}{a}$ inside the
logarithm to ensure that it is dimensionless. 
%Since we can replace the
%lattice integral of the derivatives
% with its continuum counterpart we see that equation
%(\ref{eqn:logterms}) becomes 
%\bea 
%\lim_{a \rightarrow 0} I(\mathbf{p})&=& \mathbf{L} b_C \log (p^2 +
%  \mu^2)a^2 
%\eea
%Therefore we could have extracted the leading log coefficient by
%simply taking the na\"ive continuum limit of $I_{\eta
%  \psi_d}(\mathbf{p})$.

Since all the Feynman graphs we need to evaluate are logarithmically
divergent and in one-to-one correspondence with continuum diagrams,
the resulting logarithmic divergences can all be extracted by
following a similar procedure i.e. taking the na\"ive continuum limit of
the relevant $I(\mathbf{p})$. 
\bea
\lim_{a \rightarrow 0}I_{\psi_a \chi_{gh} }^{(1)}(\mathbf{p})  &\sim&
\int \frac{d^4\mathbf{q}}{(2 \pi)^4}
\sum_{m} \frac{-i(p-q)_m}{2(q^2+\mu^2)
  ((\mathbf{p}-\mathbf{q})^2+\mu^2)}
 \left(
3\delta_{ag} \delta_{mh} 
-3\delta_{ah} \delta_{mg}\right)
f_{ABC}f_{BCD}
\nonumber \\ &\sim&
\frac{3i}{32 \pi^2}f_{ABC} f_{BCD} \left(
\delta_{ag} p_h 
-\delta_{ah} p_g \right) \log \mu a
\end{eqnarray}
\begin{eqnarray}
\lim_{a \rightarrow 0}
I_{\psi_a \chi_{de}}^{(2)}(\mathbf{p})&\sim&   
\int \frac{d^4\mathbf{q}}{(2 \pi)^4}
\sum_{c}
\frac{-i(p-q)_c}{2(q^2+\mu^2) ((\mathbf{p}-\mathbf{q})^2+\mu^2)}
 (\delta_{da} \delta_{ec}-\delta_{dc}
\delta_{ea})f_{ABC} f_{BCD}
\nonumber \\ &\sim&
 \frac{i}{32 \pi^2}f_{ABC} f_{BCD}
(\delta_{da}p_e-
\delta_{ea}p_d) \log \mu a
\end{eqnarray}
This obviously leads us to define $I_{\psi_a \chi_{de}}(\mathbf{p})=
I^{(1)}_{\psi_a \chi_{de}}(\mathbf{p})+I^{(2)}_{\psi_a
  \chi_{de}}(\mathbf{p})$ and therefore
\begin{eqnarray}
I_{\psi_a \chi_{de}}(\mathbf{p})&\sim&
 \frac{i}{8 \pi^2}f_{ABC} f_{BCD}
(\delta_{da}p_e-
\delta_{ea}p_d)
\end{eqnarray}
\begin{eqnarray}
\lim_{a \rightarrow 0}I_{\chi_{ab} \chi_{gh}} (\mathbf{p}) &\sim&
 \int \frac{d^4\mathbf{q}}{(2 \pi)^4}
\sum_{d}
\frac{i(p-q)_d}{2(q^2+\mu^2)  ((\mathbf{p}-\mathbf{q})^2+\mu^2)}
\epsilon_{abdgh}     
f_{ABC}f_{BCD} 
\nonumber \\ &&
- \left( h \leftrightarrow g \right)
\nonumber \\ &\sim&
-\frac{i}{16 \pi^2} f_{ABC} f_{BCD} \sum_d \epsilon_{abdgh} p_d \log
 \mu a
\end{eqnarray}
Note that these calculations of the $\log$ terms for the other
diagrams have also been
verified by numerical evaluation and fitting of the resulting lattice integrals.

\subsection{The Auxiliary Field Diagram}
Since the amputated divergent diagrams for the lattice $d$ propagator are log divergent we can extract the sum of these logarithmic divergences using the same tricks we used for the fermions; evaluating the diagram in the na\"ive continuum limit. The sum of all these diagrams, contracted with external $dB$ propagators, will then yield a log divergent term of the
form
\beq
C_{dd}=cf_{ACB}f_{DCB}\ln{(\mu a)}
\eeq
where $c$ is a constant to be determined by explicitly evaluating the diagrams.
However, we will argue in the next section
that there that it is not necessary to evaluate these diagrams, even
in the continuum, to determine $\alpha_3$ -- the requirement that the continuum theory preserve full
supersymmetry will automatically determine $\alpha_3$ in terms of the
other $\alpha_i$ corresponding to the fermion propagator renormalization.

\subsection{From Amputated Diagrams to Renormalized Propagators}

The leading logarithmic divergences appearing in the renormalized propagators are obtained by combining the (divergent parts of) the individual amputated diagrams we have just computed. In principle several of the amputated fermion diagrams can appear as internal bubbles when correcting a given fermion propagator. As an example consider the $\psi \eta$ diagram shown in Figure \ref{fig:full_p}. Na\"ively we see that three of our amputated diagrams contribute to the renormalization of this propagator. However we find that (at least in the case of the $\log$ divergences) the Lorentz structure of the propagators and integrals means that only the $\eta \psi$ amputated diagram contributes to the renormalization of the $\eta \psi$ propagator.

\FIGURE[t]
  {
  \includegraphics[scale=1.0]{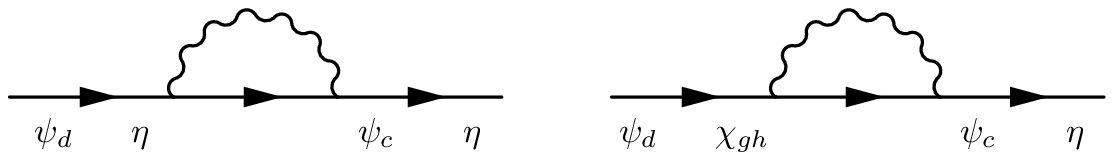}
  \caption{\label{fig:full_p}
    Full $\eta \psi$ Propagators}
  }
We demonstrate this through explicit calculation. Denoting the full diagrams by $C$ and noting that as we are dealing with only the divergent
part we can approximate the lattice propagators by their continuum analogues
we find
\begin{eqnarray}
C_{\psi_d \eta}&=& \frac{2ip_d}{p^2}
\sum_c I_{\eta \psi_c}(\mathbf{p}) \frac{2ip_c}{p^2}
+\sum_{c,g,h}
\frac{ip_g \delta_{dh} -ip_h \delta_{dg}}{p^2}
 I_{\chi_{gh} \psi_c}(\mathbf{p})
\frac{2ip_c}{p^2}
\nonumber \\ &\sim&
-\frac{2ip_d}{p^2}
\frac{i}{8 \pi^2} f_{ABC}f_{BCD}\sum_c p_c
\frac{2ip_c}{p^2} \log \mu a
\nonumber \\ &\sim&
\frac{1}{4 \pi^2} f_{ABC}f_{BCD}\frac{2ip_d}{p^2}
 \log \mu a~.
\end{eqnarray}
The second term disappears as 
\begin{eqnarray}
\sum_{c,g,h}
(p_g \delta_{dh} -p_h \delta_{dg})
( \delta_{cg} p_h - \delta_{ch}p_g)
p_c
 &=&
\sum_{c}
(p_c p_d -p^2 \delta_{cd}
-
p^2\delta_{dc} +p_c p_d
)p_c
\nonumber \\ &=&0~.
\end{eqnarray}
We can similarly show that only  $I_{\psi \chi}$ contributes to $C_{\psi \chi}$ and $I_{\chi \chi}$ to $C_{\chi \chi}$. Note, however,
that this analysis strictly only applies to the logarithmically
divergent piece in $C$. 
\begin{eqnarray}
C_{\psi_a \chi_{de}}
&=&  \frac{i}{8 \pi^2}f_{ABC} f_{BCD}\sum_{g,h,c}
\frac{ip_g \delta_{ah} -ip_h \delta_{ag}}{p^2}
(\delta_{gc}p_h-
\delta_{hc}p_g)
\frac{ip_d \delta_{ce} -ip_e \delta_{cd}}{p^2} \log \mu a
\nonumber \\ &=&
\frac{1}{4 \pi^2} f_{ABC}f_{BCD} 
\frac{ip_d \delta_{ae} -ip_e \delta_{ad}}{p^2}~\log \mu a~.
\end{eqnarray}
In calculating $C_{\chi \chi}$ we must take into account that the internal propagator in $I_{\chi \chi}$ can be a $\psi \chi$ or $\chi \psi$. This contributes another factor of 2 to $C_{\chi \chi}$.
\begin{eqnarray}
C_{\chi_{ab} \chi_{de}}&=&-\frac{i}{8 \pi^2} f_{ABC} f_{BCD}\log \mu a
\sum_{c,f,g,i,h,j,k}\epsilon_{abcfg} \frac{ip_c}{2p^2}
  \epsilon_{fgihj} p_i 
\epsilon_{hjkde} \frac{ip_k}{2p^2}
\nonumber \\ &=&
-\frac{i}{2 \pi^2} f_{ABC} f_{BCD}\log \mu a
\sum_{c,i,k} \frac{ip_c}{2p^2}
   p_i 
 \frac{ip_k}{2p^2} \left(
\delta_{ai} \epsilon_{bckde}
+\delta_{bi} \epsilon_{cakde}
+\delta_{ci} \epsilon_{abkde}
\right)
\nonumber \\ &=&
\frac{1}{4 \pi^2} f_{ABC} f_{BCD}\log \mu a
\sum_{k}  
 \frac{ip_k}{2p^2} 
 \epsilon_{abkde}~.
\end{eqnarray}
The coefficients $\alpha_i$ are now determined by the coefficient of
the propagator in the renormalized propagator amplitudes $C$. 
Explicitly we find
\beq
\alpha_i=1+b_i\ln{\mu a}\quad i=1,2,4
\eeq
where $b_i=b=\frac{g^2N}{4\pi^2}$. 
Note that we have used $f_{ABC}f_{BCD}=N \delta_{AD}$. This is
required as the
colour structure of any counterterms must match 
the tree propagators. However this is strictly only true
for $SU(N)$ as  $f_{ABC}f_{BCD}=N (\delta_{AD}-\delta_{A0}\delta_{D0})$
for $U(N)$. This does not matter in the continuum as the $U(1)$ trace piece
simply decouples from the rest of the system and can be ignored. When
doing lattice simulations we might imagine
achieving a similar result by giving the
$U(1)$ mode a large mass of the order of the cut-off which will
serve to decouple it from the $SU(N)$ modes at finite lattice
spacing. The breaking of supersymmetry in this sector may then be
removed by sending this $U(1)$ mass to zero
{\it after} taking the continuum limit.

While na\"ively one might have expected
the coefficients $b_i$ to be all different our results indicate
that in fact the
log divergent parts of $b_i$ and
hence $\alpha_i$
are actually all equal.  
This fact can be understood quite simply; to untwist 
the
continuum theory into a theory with four Majorana spinors requires that the
continuum twisted fermions exhibit a common wavefunction renormalization.
This just follows from the fact that the individual components of the
spinors mix the different twisted fermions together.
To achieve this 
requires that the corresponding renormalization constants of the
kinetic terms $\alpha_i$ should
all be equal --  just as we find. Furthermore, since the leading log
behavior of the lattice theory is the same as the continuum we 
should expect that the log divergent part of the lattice
couplings behave in the same way. Thus a single wavefunction renormalization
of the twisted lattice fermions is all that is needed to render the
renormalized theory finite. The common anomalous dimension of the
fermions in this
twisted scheme is then given by $\gamma=\frac{g^2N}{8\pi^2}$.

In the case of the $<dd>$ propagator the leading log divergent contribution
can be computed from the na\"ive continuum limit of the
corresponding continuum expression for the sum of the
$BB$ bubble diagrams given
in diagram~\ref{AllBbubbles}.
Combined with the fact that the tree level $<dB>$
propagators required on the outside of these $BB$ amputated diagrams
are the same as
the continuum to $\mathcal{O}(a)$ we find that
the log divergence in the mass
renormalisation of the $d$ field must be the same on the lattice as
in the continuum. Using this fact we can argue that the log divergent
part of $\alpha_3$ must actually be equal to that of the fermions eg
$\alpha_1$. This follows from the fact that the bosonic action for
general $\alpha_i$ can be rewritten as
\beq
\alpha_1 \Big(\cFb_{ab} \cF_{ab}\Big) + \frac{\alpha_2^2}{\alpha_3}\Big(\frac{1}{2}[\cDb_a,\cD_a]^2\Big)
\eeq
Only for $\alpha_3=\alpha_2=\alpha_1$ can this renormalized bosonic
action be untwisted to yield the conventional gauge field plus scalar action in the continuum limit. But since the continuum twisted theory 
possesses full supersymmetry this must be true. And our general arguments
then tell us the log divergence of $\alpha_3$ on the lattice must
satisfy the same property. 

To summarise; we find that the log divergent parts of the coefficients
$\alpha_i,i=1\ldots 4$ must all be equal to one loop order in the lattice
theory. This implies that a common wavefunction renormalization of both
twisted fermions and bosons is sufficient to render the renormalized
theory finite at one loop with all
fields acquiring an anomalous dimension (in this scheme) given
by $\gamma=\frac{g^2N}{8\pi^2}$.
Physically, the equality of the 
couplings $\alpha_i,i=1\ldots 4$ 
means that {\it no} logarithmic fine tuning is required
at weak coupling for the lattice theory to exhibit full supersymmetry
as the lattice spacing is sent to zero.

%%%%%%%%%%%%%%%%%%%%%%%%%%%%%%%%%%%%%%%%%%%%%%%%%%%%%%%%%%%%%%%%%
\section{Conclusion and Discussion}
%%%%%%%%%%%%%%%%%%%%%%%%%%%%%%%%%%%%%%%%%%%%%%%%%%%%%%%%%%%%%%%%%

We have examined a recently developed lattice construction for $\cN=4$, $d=4$ super Yang-Mills theory using perturbation theory. We argue that the exact symmetries of the classical lattice theory; namely gauge invariance, a single exact supersymmetry $\cQ$ and the (large) point group symmetry of the lattice strongly constrain the possible counter terms induced by quantum corrections. Indeed with one exception the only relevant counterterms
correspond to renormalizations of existing terms in the action. We furthermore show by a computation of the
effective action that the one new operator which cannot be excluded in the general analysis actually makes no appearance to all orders in perturbation theory\footnote{The calculation we have done is somewhat formal
as it ignores possible
instabilities associated with the flat directions and specifically the
$U(1)$ trace mode of the scalars. It is possible that regulating these directions
by eg introducing a mass term for the $U(1)$ mode
might modify our conclusions since supersymmetry
is broken by such terms. Notice though that at large N the dangerous
double trace mass operator is suppressed and hence the result should
certainly hold in that limit}

The renormalized action can then be written in terms of 4 coupling constants $\alpha_i$ which take the value unity in the classical lattice action. We evaluate the renormalization of these couplings at one loop using lattice perturbation theory. Three of the couplings can be computed by examining the renormalization of the three twisted fermion propagators. The final coupling is most easily read off from a one loop contribution to the propagator for a bosonic auxiliary field. The relevant propagators and vertices are derived and the amputated one loop diagrams constructed. All these diagrams possess identical logarithmic divergences of the form $\ln{\mu a}$ where $a$ is the lattice spacing and $\mu$ a mass scale introduced to regulate the small momentum behavior of the integrands. This divergence can be absorbed by a common wavefunction renormalization $Z$ of the twisted fermions and bosons.

The simplest way to understand this rather surprising result
is to realize that the coefficient of the logarithmic divergence
of some one loop diagram in the lattice theory can be extracted by 
taking a na\"ive continuum limit of the diagram, since the
log divergence comes from the small loop momentum region of the integral.
Provided that the lattice diagrams correspond one-to-one with 
equivalent continuum diagrams,\footnote{Note that this is true in the
present, twisted construction and that it is not true in other,
more na\"ive constructions.  There, one would have ${\cal O}(a)$
multigluon vertices that correct the fermion self energy
through a tadpole.  This lattice diagram does not
occur in the perturbation theory we have described above.}
and that all lattice propagators and vertices reduce to their continuum counterparts for small momenta, this means that the log divergences of the lattice theory are equal to the same divergences in the continuum theory\footnote{Clearly both continuum and lattice perturbation theories must employ the same scheme for this to be true. In this light we would note that it is possible by addition of sufficient auxiliary fields to arrange for $Z=1$ for all fields in continuum $\cN=4$ SYM. However such a superfield approach is not possible for our twisted construction. This does not spoil the finiteness of the theory which depends only on the vanishing of the beta function which is true in both twisted and supergraph schemes.}. Furthermore, since the twisted continuum theory is equivalent to the usual $\cN=4$ theory in flat space it must possess the full $\cQ=16$ supersymmetry. This fact ensures that all divergences present in the twisted continuum fermion self energies must be equal - which is indeed what we find. And this structure is necessarily inherited by the log
divergent parts of the lattice theory at one loop.  This is what leads to our main result; that only a one
time tuning of the
finite parts of the wavefunction renormalization needs to be performed at one
loop in order to restore the full supersymmetry.

This similarity between the divergence structure of the lattice theory and the continuum theory is strongly suggestive that the beta function of the lattice theory will also vanish at weak coupling. We leave a proof of this (and a computation of the finite parts of the diagrams) to a later paper and here merely give heuristic arguments as to why this result may indeed hold. First, note that the calculation of the beta function requires the evaluation of one loop vertex diagrams in the lattice theory. Preliminary
calculations suggest that the
set of relevant lattice vertex diagrams correspond one to one to continuum vertex
diagrams and remain only logarithmically divergent.
They may thus be evaluated in the continuum theory. 
The coefficient of this log divergence is then combined with the wavefunction renormalizations determined above
to yield the one loop beta function in the usual manner. 
However we already know the result of this computation for the continuum theory; the beta function vanishes. We hence expect a similar result to hold at one loop in the lattice theory. Thus for weak coupling we expect the lattice theory to possess a line of fixed points parametrized by the bare coupling constant just as for the continuum theory. However, our calculations do not reveal whether this feature survives in the lattice theory
to strong coupling. At two or more loops the divergences of the lattice Feynman diagrams will not be equal to the those of the continuum theory and hence we cannot use the latter to infer the divergence structure of the lattice theory.
To understand how to take the continuum limit in this regime will then require a mixture of two loop and numerical calculations.

%%%%%%%%%%%%%%%%%%%%%%%%%%%%%%%%%%%%%%%%%%%%%%%%%%%%%%%%%%%%%%%%%
\acknowledgments
%%%%%%%%%%%%%%%%%%%%%%%%%%%%%%%%%%%%%%%%%%%%%%%%%%%%%%%%%%%%%%%%%

S.C. and A.J. are supported in part by the U.S. Department of Energy grant under Contract No. DE-FG02-85ER40237. J.G.'s work is supported in part by the U.S. Department of Energy grant under Contract No. DE-FG02-08ER41575 as well as Rensselaer faculty development funds. R.W. is supported by an STFC studentship and would like to thank Ron Horgan for useful conversations. S.C. acknowledges useful discussions with Poul Damgaard, David B. Kaplan and Mithat \"{U}nsal.

\appendix
%%%%%%%%%%%%%%%%%%%%%%%%%%%%%%%%%%%%%%%%%%%%%%%%%%%%%%%%%%%%%%%%%

\section{Simplification of the One Loop Diagrams}

This section shows how the diagrams in the paper are simplified. For the easy evaluation of the diagram we use the following identities.
\beq
\sum_{B,C} \lambda_{ABC} \lambda_{BCD} = \sum_{B,C} d_{ABC} d_{BCD} - f_{ABC} f_{BCD}~.
\eeq
\beq
\overline{\lambda}_{ABC} = \lambda_{ACB}~.
\eeq
These relations imply
\bea
(A \lambda_{ABC} - B \overline{\lambda}_{ABC})(C \lambda_{BCD} - D \overline{\lambda}_{BCD})
&=& d_{ABC}d_{BCD} (AC-AD-BC+BD)\nn \\
&&- f_{ABC} f_{BCD}(AC +AD +BC +BD)~.~~~~
\eea

Starting with $I_{\eta \psi}(\mathbf{p})$:
\bea
I_{\eta \psi_d}(\mathbf{p}) &=& \int \frac{d^4 \mathbf{q}}{(2 \pi)^4} \sum_{a,b,c} \sum_{BC}
   \left[ \frac{1}{ 16\widehat{(\mathbf{p}-\mathbf{q})}^2 
\widehat{\mathbf{q}}^2} 
\right] 
[\lambda_{ABC} - \overline{\lambda}_{ABC}e^{ip_a}] 
 \left[[\overline{\lambda}_{BCD}e^{-ip_a} -
  \lambda_{BCD}e^{iq_d}]\right]
\nonumber \\ &&
\left((1-e^{i(p-q)_b})\delta_{ca}(\delta_{ba}\delta_{cd} -
   \delta_{bd}\delta_{ca})-(1 - e^{i(p-q)_c})(\delta_{ba}\delta_{cd} -
   \delta_{bd}\delta_{ca}) \delta_{ba} \right)\nn \\
   &=&\int \frac{d^4 \mathbf{q}}{(2 \pi)^4} \sum_{a,b} \sum_{BC}
   \left[ \frac{1}{ 8\widehat{(\mathbf{p}-\mathbf{q})}^2 
\widehat{\mathbf{q}}^2} 
\right] 
\nonumber \\ &&
\left[
d_{ABC} d_{BCD} (e^{-ip_a} -e^{iq_d} -1 + e^{ip_a +iq_d})
+f_{ABC} f_{BCD} (e^{-ip_a} +e^{iq_d} +1 + e^{ip_a +iq_d})
\right]
\nonumber \\ &&
(1-e^{i(p-q)_b})(\delta_{ba}\delta_{ad} -
   \delta_{bd} \delta_{aa}) \nonumber 
\eea

That is
\bea
I_{\eta \psi_d}(\mathbf{p})
&=&
\int \frac{d^4 \mathbf{q}}{(2 \pi)^4} \sum_{a} \sum_{BC}
   \left[ \frac{1}{8 \widehat{(\mathbf{p}-\mathbf{q})}^2 
\widehat{\mathbf{q}}^2} 
\right] 
\bigg[ \big[
d_{ABC} d_{BCD} (e^{-ip_a} -e^{iq_d} -1 + e^{ip_a +iq_d})
\nonumber \\ &&
+f_{ABC} f_{BCD} (e^{-ip_a} +e^{iq_d} +1 + e^{ip_a +iq_d})
\big] (1-e^{i(p-q)_a}) \delta_{ad}
\nonumber \\ &&
-\big[
d_{ABC} d_{BCD} (e^{-ip_a} -e^{iq_d} -1 + e^{ip_a +iq_d})
\nonumber \\ &&
+f_{ABC} f_{BCD} (e^{-ip_a} +e^{iq_d} +1 + e^{ip_a +iq_d})
\big]
(1-e^{i(p-q)_d})   
\bigg]
\nonumber \\
&=&
\int \frac{d^4 \mathbf{q}}{(2 \pi)^4}  \sum_{BC}
   \left[ \frac{1}{8 \widehat{(\mathbf{p}-\mathbf{q})}^2 
\widehat{\mathbf{q}}^2} (1-e^{i(p-q)_d}) 
\right] 
\bigg[ \big[
d_{ABC} d_{BCD} (e^{-ip_d} -e^{iq_d} -1 + e^{i(p+q)_d})
\nonumber \\ &&
+f_{ABC} f_{BCD} (e^{-ip_d} +e^{iq_d} +1 + e^{i(p+q)_d})
\big] 
\nonumber \\ &&
-\sum_a \big[
d_{ABC} d_{BCD} (e^{-ip_a} -e^{iq_d} -1 + e^{ip_a +iq_d})
\nonumber \\ &&
+f_{ABC} f_{BCD} (e^{-ip_a} +e^{iq_d} +1 + e^{ip_a +iq_d})
\big]
\bigg]
\nonumber \\
&=&
\int \frac{d^4 \mathbf{q}}{(2 \pi)^4}  \sum_{BC}
   \left[ \frac{1}{8 \widehat{(\mathbf{p}-\mathbf{q})}^2 
\widehat{\mathbf{q}}^2} (1-e^{i(p-q)_d}) 
\right] 
\bigg[
-\sum_{a \neq d} \big[
d_{ABC} d_{BCD} (e^{-ip_a} -e^{iq_d} -1 + e^{ip_a +iq_d})
\nonumber \\ &&
+f_{ABC} f_{BCD} (e^{-ip_a} +e^{iq_d} +1 + e^{ip_a +iq_d})
\big]
\bigg]
\end{eqnarray}
Now $I_{\psi_a \chi_{gh}}^{(1)}$:
\bea
I_{\psi_a \chi_{gh} }^{(1)}(\mathbf{p})  &=&
\int \frac{d^4\mathbf{q}}{(2 \pi)^4}
\sum_{b,c,d,e,m,f} \sum_{B,C} \frac{(-1)}{64\widehat{\mathbf{q}}^2
  \widehat{(\mathbf{p}-\mathbf{q})}^2}
\epsilon_{bcmef} \epsilon_{ghdef} e^{i(p-q)_{(e+f)}}
\nonumber \\ &&(e^{i(p-q)_m}-1)
(\delta_{bd} \delta_{ca} - \delta_{ba} \delta_{cd})
\nonumber \\ &&
\times \left( \lambda_{ABC} e^{ip_d} - \overline{\lambda}_{ABC}
e^{-iq_a} \right)
\bigg(e^{-ip_{(d+e+f)}}
\left( \overline{\lambda}_{BCD} e^{iq_{(e+f)}} - \lambda_{BCD}
e^{i(p-q)_{d}} \right)
\nonumber \\ &&
-e^{i(p-q)_{(d+g+h)}}
\left( \lambda_{BCD} e^{iq_{(g+h)}} - \overline{\lambda}_{BCD}
e^{-ip_{d}} \right) 
\bigg)
\nonumber
\eea
That is
\bea
I_{\psi_a \chi_{gh} }^{(1)}(\mathbf{p})&=&
\int \frac{d^4\mathbf{q}}{(2 \pi)^4}
\sum_{d,e,m,f} \sum_{B,C} \frac{1}{16\widehat{\mathbf{q}}^2
  \widehat{(\mathbf{p}-\mathbf{q})}^2}
\epsilon_{admef} \epsilon_{ghdef}
(e^{i(p-q)_m}-1)
\nonumber \\ &&
\times 
\big(-d_{ABC}d_{BCD} 
(e^{i(p_d+q_{(g+h)})}
-1-e^{iq_{(g+h-a)}}
+e^{-i(p_d+q_a)})
\nonumber \\ &&+
f_{ABC}f_{BCD}(e^{i(p_d+q_{(g+h)})}
+1+e^{iq_{(g+h-a)}}
+e^{-i(p_d+q_a)})
\big)
\nonumber \\ &=&
\int \frac{d^4\mathbf{q}}{(2 \pi)^4}
\sum_{d,m} \sum_{B,C} \frac{1}{8\widehat{\mathbf{q}}^2
  \widehat{(\mathbf{p}-\mathbf{q})}^2}
(e^{i(p-q)_m}-1)\nonumber \\ &&
\times \left(
\delta_{ag} \delta_{mh} 
+\delta_{ah} \delta_{md} \delta_{dg}
+\delta_{ad} \delta_{mg} \delta_{dh}
-\delta_{ah} \delta_{mg} 
-\delta_{ag} \delta_{md} \delta_{dh}
-\delta_{ad} \delta_{mh} \delta_{dg}
\right)
\nonumber \\ &&
\times 
\big(d_{ABC}d_{BCD} 
(e^{i(p_d+q_{(g+h)})}
-1-e^{iq_{(g+h-a)}}
+e^{-i(p_d+q_a)})
\nonumber \\ &&-
f_{ABC}f_{BCD}(e^{i(p_d+q_{(g+h)})}
+1+e^{iq_{(g+h-a)}}
+e^{-i(p_d+q_a)})
\big)
\end{eqnarray}
Looking at the second $\psi \chi$ diagram we have:
\begin{eqnarray}
I_{\psi_a \chi_{de}}^{(2)}(\mathbf{p})&=&   
\int \frac{d^4\mathbf{q}}{(2 \pi)^4}
\sum_{b,c,B,C}
\frac{1}{8\widehat{\mathbf{q}}^2  \widehat{(\mathbf{p}-\mathbf{q})}^2}
(e^{i(p-q)_c}-1) \delta_{ab}(\delta_{db} \delta_{ec}-\delta_{dc}
\delta_{eb}) \nonumber\\ &&
\times(\lambda_{ABC} e^{-i(p -q)_a}-\overline{\lambda}_{ABC})
(\lambda_{BCD} e^{iq_c} - \overline{\lambda}_{BCD} e^{i(p-q)_b})
\nonumber \\
&=&   
\int \frac{d^4\mathbf{q}}{(2 \pi)^4}
\sum_{c,B,C}
\frac{1}{8\widehat{\mathbf{q}}^2  \widehat{(\mathbf{p}-\mathbf{q})}^2}
(e^{i(p-q)_c}-1) (\delta_{da} \delta_{ec}-\delta_{dc}
\delta_{ea}) \nonumber\\ &&
\times \bigg(
d_{ABC} d_{BCD}(e^{i(q_c-(p-q)_a)} - e^{iq_c} -1+e^{i(p-q)_a} )
\nonumber \\ &&-
f_{ABC} f_{BCD}(e^{i(q_c-(p-q)_a)} + e^{iq_c}+1 +e^{i(p-q)_a} )
\bigg)
\end{eqnarray}
Now looking at $I_{\chi_{ab} \chi_{gh}}$:
\begin{eqnarray}
I_{\chi_{ab} \chi_{gh}} (\mathbf{p}) &=& \int \frac{d^4\mathbf{q}}{(2 \pi)^4}
\sum_{c,d,e,f,B,C}
\frac{1}{32\widehat{\mathbf{q}}^2  \widehat{(\mathbf{p}-\mathbf{q})}^2}
 \epsilon_{abcde} (\delta_{gc} \delta_{hf} -
\delta_{gf} \delta_{hc})  \nonumber \\ &&
\times
\left( (e^{-i(p-q)_d}-1) \delta_{ef} - (e^{-i(p-q)_e} -1) \delta_{df}
\right) \left( \lambda_{BCD} e^{iq_f} - \overline{\lambda}_{BCD}
e^{i(p-q)_{c}} \right)
\nonumber \\ &&
\times \Big(
e^{-ik_{(a+b+c)}}\left( \overline{\lambda}_{ABC} e^{ip_c} - \lambda_{ABC}
e^{-iq_{(a+b)}} \right)
\nonumber \\ &&
-e^{ip_{(c+d+e)}}\left( \lambda_{ABC} e^{-i(p-q)_c} - \overline{\lambda}_{ABC}
e^{-iq_{(d+e)}} \right) \Big)
\nn
\eea

That is 
\bea
I_{\chi_{ab} \chi_{gh}} (\mathbf{p}) &=&
 \int \frac{d^4\mathbf{q}}{(2 \pi)^4}
\sum_{d,e,B,C}
\frac{-1}{16\widehat{\mathbf{q}}^2  \widehat{(\mathbf{p}-\mathbf{q})}^2}
e^{ip_{(g+d+e)}} \epsilon_{abgde}  \nonumber \\ &&
\times
\left( (e^{-i(p-q)_d}-1) \delta_{eh} - (e^{-i(p-q)_e} -1) \delta_{dh}
\right) 
\nonumber \\ &&
\times \left( \lambda_{ABC} e^{-i(p-q)_g} - \overline{\lambda}_{ABC}
e^{-iq_{(d+e)}} \right)
\left( \lambda_{BCD} e^{iq_h} - \overline{\lambda}_{BCD}
e^{i(p-q)_{g}} \right)
\nonumber \\ &&
- \left( h \leftrightarrow g \right)
 \nonumber \\ &=&
 \int \frac{d^4\mathbf{q}}{(2 \pi)^4}
\sum_{d,B,C}
\frac{1}{8\widehat{\mathbf{q}}^2  \widehat{(\mathbf{p}-\mathbf{q})}^2}
e^{ip_{(g+d+h)}} \epsilon_{abdgh}   (e^{-i(p-q)_d}-1)   
\nonumber \\ &&
\times \big( d_{ABC}d_{BCD} (e^{-i(p_g-q_{(g+h)})} -1 -e^{-iq_d} + e^{i(p_g
  -q_{(g+d+h)})})
\nonumber \\ &&- 
f_{ABC}f_{BCD}(e^{-i(p_g-q_{(g+h)})} +1 +e^{-iq_d} + e^{i(p_g
  -q_{(g+d+h)})}) \big)
\nonumber \\ &&
- \left( h \leftrightarrow g \right)
\end{eqnarray}
(Note that we also need to take into account the diagram where the internal $\psi \chi$ is flipped. It is the same as what we have but with $a
\leftrightarrow g$, $b \leftrightarrow h$ and $\mathbf{p} \leftrightarrow -\mathbf{p}$. We may for convenience take $\mathbf{q} \leftrightarrow -\mathbf{q}$. We pick up an additional minus sign in the $f_{ABC} f_{BCD}$ term due to the differing order of the group factors.)

%%%%%%%%%%%%%%%%%%%%%%%%%%%%%%%%%%%%%%%%%%%%%%%%%%%%%%%%%%%%%%%%%
\section{Vanishing of One Loop Fermion Propagator at Zero Momentum}
%%%%%%%%%%%%%%%%%%%%%%%%%%%%%%%%%%%%%%%%%%%%%%%%%%%%%%%%%%%%%%%%%
Starting with the first diagram and using the simplified forms of the
integrals derived in Appendix A (assuming an IR regulator)
we have:
\begin{eqnarray}
I_{\eta \psi_d}(\mathbf{0})
&=&
\int \frac{d^4 \mathbf{q}}{(2 \pi)^4}  \sum_{BC}
   \left[ \frac{1}{8 \widehat{\mathbf{q}}^2 
\widehat{\mathbf{q}}^2} (1-e^{-iq_d}) 
\right] 
\bigg[
-2\sum_{a \neq d} 
f_{ABC} f_{BCD} (1 +e^{iq_d})
\bigg]
\nonumber \\ &=& 0
\end{eqnarray}
as $(1-e^{-iq_d})(1+e^{iq_d})=2i \sin q_d$ and then the integrand
is the combination of an odd and an even function.
Next we calculate:
\begin{eqnarray}
I_{\psi_a \chi_{gh}}^{(1)}(\mathbf{0})
 &=&
\int \frac{d^4\mathbf{q}}{(2 \pi)^4}
\sum_{d,m} \sum_{B,C} \frac{1}{8\widehat{\mathbf{q}}^2
  \widehat{\mathbf{q}}^2}
(e^{-iq_m}-1)\nonumber \\ &&
\times \left(
\delta_{ag} \delta_{mh} 
+\delta_{ah} \delta_{md} \delta_{dg}
+\delta_{ad} \delta_{mg} \delta_{dh}
-\delta_{ah} \delta_{mg} 
-\delta_{ag} \delta_{md} \delta_{dh}
-\delta_{ad} \delta_{mh} \delta_{dg}
\right)
\nonumber \\ &&
\times 
\big(d_{ABC}d_{BCD} 
(e^{iq_{(g+h)}}
-1-e^{iq_{(g+h-a)}}
+e^{-iq_a})
\nonumber \\ &&-
f_{ABC}f_{BCD}(e^{iq_{(g+h)}}
+1+e^{iq_{(g+h-a)}}
+e^{-iq_a})
\big)
\nonumber \\ &=&
\int \frac{d^4\mathbf{q}}{(2 \pi)^4}
\sum_{m} \sum_{B,C} \frac{1}{8\widehat{\mathbf{q}}^2
  \widehat{\mathbf{q}}^2}
(e^{-iq_m}-1)\nonumber  \left(
\delta_{ah} \delta_{mg} -
\delta_{ag} \delta_{mh} \right)
\nonumber \\ &&
\times 
\big(d_{ABC}d_{BCD} 
(e^{iq_{(g+h)}}
-1-e^{iq_{(g+h-a)}}
+e^{-iq_a})
\nonumber \\ &&-
f_{ABC}f_{BCD}(e^{iq_{(g+h)}}
+1+e^{iq_{(g+h-a)}}
+e^{-iq_a})
\big)
\end{eqnarray}
Then we can use the fact that if $a\neq g$ and $a \neq h$ then the
expression disappears. If $a=g=h$ again the expression disappears. So
assuming $a=h$ and $a\neq g$ we get:
\begin{eqnarray}
I_{\psi_a \chi_{gh}}^{(1)}(\mathbf{0})
 &=&
\int \frac{d^4\mathbf{q}}{(2 \pi)^4}
 \sum_{B,C} \frac{1}{8\widehat{\mathbf{q}}^2
  \widehat{\mathbf{q}}^2}
(e^{-iq_g}-1)\nonumber 
\big(d_{ABC}d_{BCD} 
(e^{iq_{(g+a)}}
-1-e^{iq_{g}}
+e^{-iq_a})
\nonumber \\&&
-
f_{ABC}f_{BCD}(e^{iq_{(g+a)}}
+1+e^{iq_{g}}
+e^{-iq_a})
\big)
\nonumber \\ 
&=&\int \frac{d^4\mathbf{q}}{(2 \pi)^4}
 \sum_{B,C} \frac{2i}{8\widehat{\mathbf{q}}^2
  \widehat{\mathbf{q}}^2}
\big(d_{ABC}d_{BCD} 
(\sin q_a + \sin q_g -\sin q_{(a+g)}
)
\nonumber \\&&
-
f_{ABC}f_{BCD}(
\sin q_a - \sin q_g -\sin q_{(a+g)}
)
\big)\nonumber \\ &=&0
\end{eqnarray}
which vanishes term by term. We then move onto the second $\psi \chi$ diagram.
\begin{eqnarray}
I_{\psi_a \chi_{de}}^{(2)}(\mathbf{0}) 
&=&   
\int \frac{d^4\mathbf{q}}{(2 \pi)^4}
\sum_{c,B,C}
\frac{1}{8\widehat{\mathbf{q}}^2  \widehat{\mathbf{q}}^2}
(e^{-iq_c}-1) (\delta_{da} \delta_{ec}-\delta_{dc}
\delta_{ea}) \nonumber\\ &&
\times \bigg(
d_{ABC} d_{BCD}(e^{iq_{(c+a)}} - e^{iq_c} -1+e^{-iq_a} )
\nonumber \\ &&-
f_{ABC} f_{BCD}(e^{iq_{(c+a)}} + e^{iq_c} +1+e^{-iq_a} )
\bigg)
\end{eqnarray}
In a similar way to the previous diagram if $a \neq d$ and $a \neq e$
then the diagram vanishes. If $a=d=e$ it also vanishes so we only
need to deal with the case $a=d$, $a \neq e$: 
\begin{eqnarray}
I_{\psi_a \chi_{de}}^{(2)}(\mathbf{0}) 
&=&   
\int \frac{d^4\mathbf{q}}{(2 \pi)^4}
\sum_{B,C}
\frac{1}{8\widehat{\mathbf{q}}^2  \widehat{\mathbf{q}}^2}
(e^{-iq_e}-1)   \bigg(
d_{ABC} d_{BCD}(e^{iq_{(e+a)}} - e^{iq_e} -1+e^{-iq_a} )
\nonumber \\ &&-
f_{ABC} f_{BCD}(e^{iq_{(e+a)}} + e^{iq_e} +1+e^{-iq_a} )
\bigg)
\nonumber \\ &=&
\int \frac{d^4\mathbf{q}}{(2 \pi)^4}
\sum_{B,C}
\frac{2i}{8\widehat{\mathbf{q}}^2  \widehat{\mathbf{q}}^2}
(e^{-iq_e}-1)   \bigg(
d_{ABC} d_{BCD}(\sin q_a +\sin q_{(a+e)}+\sin q_e )
\nonumber \\ &&-
f_{ABC} f_{BCD}(\sin q_a +\sin q_{(a+e)}-\sin q_e )
\bigg)
\nonumber \\ &=&0
\end{eqnarray}
which again vanishes term by term. Finally we show that $I_{\chi
  \chi}(\mathbf{0})=0$.
\begin{eqnarray}
I_{\chi_{ab} \chi_{gh}} (\mathbf{0})  &=&
 \int \frac{d^4\mathbf{q}}{(2 \pi)^4}
\sum_{d,B,C}
\frac{1}{8\widehat{\mathbf{q}}^2  \widehat{\mathbf{q}}^2}
 \epsilon_{abdgh}   (e^{iq_d}-1)   
\nonumber \\ &&
\times \big( d_{ABC}d_{BCD} (e^{iq_{(g+h)}} -1 -e^{-iq_d} + e^{-iq_{(g+d+h)}})
\nonumber \\ &&- 
f_{ABC}f_{BCD} (e^{iq_{(g+h)}} +1 +e^{-iq_d} + e^{-iq_{(g+d+h)}})
 \big)
\nonumber \\ &&
- \left( h \leftrightarrow g \right)
\nonumber \\ &=&
 \int \frac{d^4\mathbf{q}}{(2 \pi)^4}
\sum_{d,B,C}
\frac{2i}{8\widehat{\mathbf{q}}^2  \widehat{\mathbf{q}}^2}
 \epsilon_{abdgh}   
\nonumber \\ &&
\times \big( d_{ABC}d_{BCD} (
\sin q_{(d+g+h)} - \sin q_d -\sin q_{(g+h)}
)
\nonumber \\ &&- 
f_{ABC}f_{BCD}  (
\sin q_{(d+g+h)} + \sin q_d -\sin q_{(g+h)}
)
 \big)
\nonumber \\&=&0~.
\end{eqnarray}

%%%%%%%%%%%%%%%%%%%%%%%%%%%%%%%%%%%%%%%%%%%%%%%%%%%%%%%%%%%%%%%%%
\section{Coupling Constant Independence in $\cN=4$ SYM}
%%%%%%%%%%%%%%%%%%%%%%%%%%%%%%%%%%%%%%%%%%%%%%%%%%%%%%%%%%%%%%%%%

The continuum twisted $\cN = 4$ SYM discussed in this paper possesses a privileged set of operators whose expectation values can be shown to be independent of the background metric and hence topological. The condition for this to be true is that the operator be  annihilated by the charge $\cQ$. In addition the expectation values of these operators can be shown to be independent of the coupling constant. As we will see this property remains true in the lattice theory and provides powerful constraints on the renormalization of such operators. To see this result consider the twisted lattice action which is the sum of $\cQ$-exact and $\cQ$-closed terms. The coupling constant dependence of the $\cQ$-closed term can be removed, without disturbing the $\cQ$ BRST transformation, by rescaling the fields in appropriate ways. We show this below. 

The twisted action is
\bea
S &=& \frac{1}{g^2} S_{exact} + \frac{1}{g^2} S_{closed}\nn \\
&=& \int \Tr  \Big\{\frac{1}{g^2} \Big(\cFb_{ab}\cF_{ab} + \frac{1}{2}[\cDb_a, \cD_a]^2 - \chi_{ab}\cD_{[a}\psi_{b]} - \eta\cDb_{a}\psi_{a}\Big) \nn \\
&& - \frac{1}{g^2}  \Big(\frac{1}{2}\epsilon_{abcde}\chi_{ab}\cDb_{c}\chi_{de}\Big)\Big\}~.
\eea
A simple rescaling of the fields 
\bea
&&\chi_{ab} \rightarrow \chi_{ab}/g,~~~ \psi_{a} \rightarrow g \psi_{a},~~~\eta \rightarrow \eta/g~,
\eea
gives the action 
\bea
\label{eq:action-g-scaled}
S &=& \frac{1}{g^2} \int \Tr \Big(\cFb_{ab}\cF_{ab} + \frac{1}{2}[\cDb_a, \cD_a]^2 - \chi_{ab}\cD_{[a}\psi_{b]} - \eta\cDb_{a}\psi_{a}\Big) \nn \\
&&- \frac{1}{2} \int \Tr \epsilon_{abcde}\chi_{ab}\cDb_{c}\chi_{de}\nn \\
&=& \frac{1}{g^2} S_{exact} +  S_{closed}
\eea

Calling $\beta=\frac{1}{g^2}$ and writing the action as $S = \cQ\Lambda+S_{\rm closed}$ the expression for the expectation value of a $\cQ$-invariant operator $\cO$ becomes
\beq
\langle \cO \rangle_{\beta} = \frac{1}{Z}\int \cO e^{-(\beta\cQ \Lambda + S_{closed})},~~~Z = \int e^{-(\beta\cQ \Lambda + S_{closed})}~.
\eeq

Differentiating this expression with respect to $\beta$ leads to 
\bea
\frac{\partial}{\partial \beta} \langle \cO \rangle_{\beta} &=& \langle \cQ \Lambda \rangle_{\beta} \langle \cO \rangle_{\beta} - \langle \cO \cQ \Lambda \rangle_{\beta} \nn \\
&=&  \langle \cQ \Lambda \rangle_{\beta} \langle \cO \rangle_{\beta} - \langle \cQ (\cO \Lambda) \rangle_{\beta} \nn \\
&=& 0~,
\eea
where we have used the fact that as long as the BRST symmetry is not broken spontaneously, the expectation value of the $\cQ$ variation
of some operator vanishes.
Thus expectation values of $\cQ$-invariant observables are independent of $\beta$ and hence can be computed exactly in the semi-classical limit $\beta \rightarrow \infty$. In this limit we need only do one loop calculations around the classical vacua.

\end{document}